\documentclass[reprint, amsmath,amssymb, prb,superscriptaddress]{revtex4-2}

\usepackage{graphicx}
\usepackage{dcolumn}
\usepackage{bm}
\usepackage[usenames,dvipsnames]{xcolor}
\usepackage{hyperref}
\usepackage[inline]{enumitem}
\usepackage{bbold}
\usepackage{soul}
\usepackage{placeins}

\begin{document}

\title{Majorana bound states in germanium Josephson junctions via phase control}

\author{Melina Luethi}
\email{melina.luethi@unibas.ch}
\affiliation{Department of Physics, University of Basel, Klingelbergstrasse 82, CH-4056 Basel, Switzerland}

\author{Henry F. Legg}
\affiliation{Department of Physics, University of Basel, Klingelbergstrasse 82, CH-4056 Basel, Switzerland}

\author{Katharina Laubscher}
\affiliation{Condensed Matter Theory Center and Joint Quantum Institute, Department of Physics, University of Maryland, College Park, Maryland 20742, USA}

\author{Daniel Loss}
\affiliation{Department of Physics, University of Basel, Klingelbergstrasse 82, CH-4056 Basel, Switzerland}

\author{Jelena Klinovaja}
\affiliation{Department of Physics, University of Basel, Klingelbergstrasse 82, CH-4056 Basel, Switzerland}

\date{\today}

\begin{abstract}
We consider superconductor-normal-superconductor-normal-superconductor (SNSNS) planar Josephson junctions in hole systems with spin-orbit interaction that is cubic in momentum (CSOI).
Using only the superconducting phase difference, we find parameter regimes where junctions of experimentally achievable transparency can enter a topological superconducting phase with Majorana bound states (MBSs) at the junction ends.
In planar germanium  heterostructures CSOI can be the dominant form of SOI and extremely strong. We show analytically and numerically that, within experimental regimes, our results provide an achievable roadmap for a new MBS platform with low disorder, minimal magnetic fields, and very strong spin-orbit interaction, overcoming many of the key deficiencies that have so far prevented the conclusive observation of MBSs. 
\end{abstract}

\maketitle

\section{Introduction}
Majorana bound states (MBSs)~\cite{kitaev2001unpaired, leijnse2012introduction, qi2011topological, beenakker2013search, sato2016majorana, pawlak2019majorana, laubscher2021majorana} 
hold promise
for topological quantum computing~\cite{kitaev2003fault, nayak2008non, elliot2015colloquium}.
However, despite enormous
effort there has been no conclusive observation of MBSs so far. Key reasons postulated for 
 this are: (1)~the level of disorder, 
 which can result in spurious signals
 that mimic 
MBSs~\cite{kells2012near, lee2012zero, rainis2013towards, roy2013topologically, ptok2017controlling, liu2017andreev, moore2018two, moore2018quantized, reeg2018zero, vuik2019reproducing, stanescu2019robust, woods2019zero, chen2019ubiquitous, awoga2019supercurrent, prada2020andreev, yu2021non, sarma2021disorder, valentini2021nontopological, hess2021local,hess2022trivial}; 
(2)~the small energy scales involved, especially due to the metalization of a semiconductor by a superconductor~\cite{reeg2017transport, reeg2017finite, reeg2018metallization, reeg2018proximity, antipov2018effects, woods2018effective, kiendl2019proximity, winkler2019unified, awoga2022robust}; 
and (3)~many protocols require large magnetic fields~\cite{sato2009non, sau2010generic, alicea2010majorana, lutchyn2010majorana, oreg2010helical, sau2010non, black2011majorana, prada2012transport, black2012edge, klinovaja2012helical, sau2013topological, dutreix2014majorana, hell2017two, pientka2017topological} 
that 
are detrimental to superconductors.

Germanium (Ge) has been one of the most 
used semiconductors
since the early days of 
electronics. 
The continual interest in Ge~\cite{wagner1989observation, murakami1991strain, xie1993very, irisawa2003hole, rossner2003effective, rossner2004scattering, shah2010reverse, kloeffel2011strong, dobbie2012ultra, foronda2014weak, mironov2016fractional, lodari2019light, scappucci2020germanium, lodari2021low} has resulted in an extremely high quality material~\cite{mizokuchi2018ballistic, scappucci2020germanium}, with 
ultralong 
mean free paths (MFPs) of up to $30$~$\mu$m~\cite{myronov2023holes}.
Also, two-dimensional hole gases (2DHG) in 
Ge 
have become prominent 
platforms
for quantum information processing~\cite{hu2007ge, hu2012hole, ares2013sige, watzinger2018germanium, li2018coupling, sammak2019shallow, hendrickx2021four, jirovec2021singlet, wang2021optimal}.
Recently, there have also been significant advances in fabricating 
hybrid 
superconductor-Ge devices~\cite{xiang2006ge, hendrickx2018gate, ridderbos2018josephson, hendrickx2019ballistic, vigneau2019germanium, aggarwal2021enhancement, tosato2023hard}.
Finally, a most attractive feature of Ge
 is the large spin-orbit interaction (SOI)~\cite{moriya2014cubic, morrison2014observation, failla2015narrow, mizokuchi2017hole, chou2018weak}, enabling, e.g., ultrafast qubit operations~\cite{scappucci2020germanium, hao2010strong, hendrickx2020single, hendrickx2020fast, froning2021ultrafast, bosco2021squeezed, wang2022ultrafast}. 
Most notably,   cubic SOI  (CSOI) is very strong in  Ge 2DHGs~\cite{marcellina2017spin, terrazos2021theory, michal2021longitudinal}
\footnote{
	Furthermore, metalization of a semiconductor by a superconductor usually decreases the SOI~\cite{reeg2018proximity, reeg2018metallization}; however, it was recently shown that metalization also modifies the wave functions in Ge such that the SOI  can actually increase~\cite{adelsberger2023}.
}
and 
results in spin-split Fermi surfaces with large mismatches in velocities,
playing a central role in the following.

\begin{figure}
	\centering
	\includegraphics[width=\columnwidth]{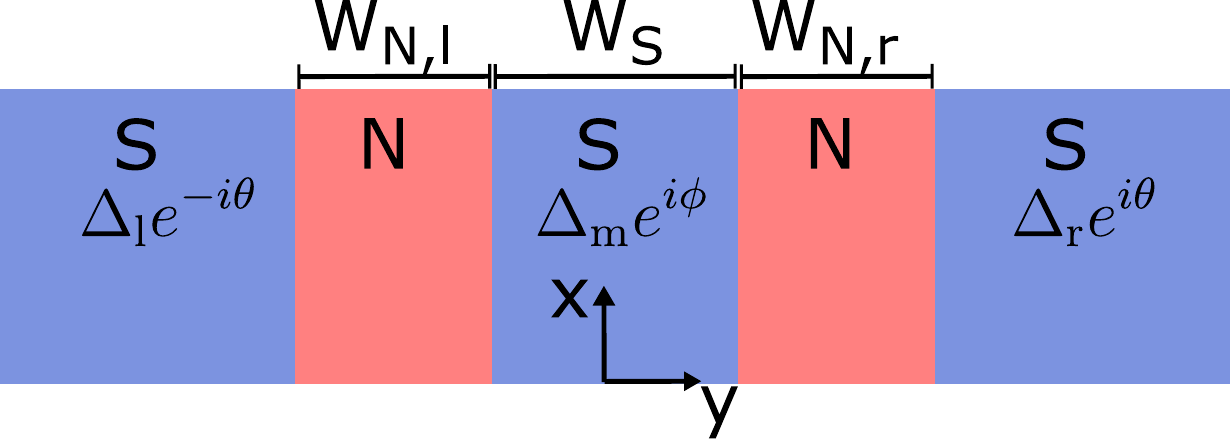}
	\caption{
		An SNSNS junction consists of three sections with proximity-induced superconductivity (blue) and two normal conducting sections (red) of width $W_{\mathrm{N,l}}$ and $W_{\mathrm{N,r}}$. The proximity-induced superconducting gaps are $\Delta_{\mathrm{l}}$, $\Delta_{\mathrm{m}}$, and $\Delta_{\mathrm{r}}$,  and the corresponding superconducting phases are $-\theta$, $\phi$, and $\theta$. The middle superconductor has a width $W_{\mathrm{S}}$.
		The two outer superconductors have widths $W_\textrm{S,l}$ and $W_\textrm{S,r}$. In the analytical calculations, $W_\textrm{S,l}$ and $W_\textrm{S,r}$ are infinite, whereas they are finite for the numerical calculations. 
		}
	\label{fig:setup}
\end{figure}

Despite ultralong MFPs and strong SOI, the  small in-plane $g$ factor of Ge ($|g|\lesssim 1.5$)~\cite{watzinger2016heavy, lu2017effective, hendrickx2018gate, hofmann2019assessing, hendrickx2020fast, scappucci2020germanium, gao2020site} 
is 
a considerable 
obstacle to
realize MBSs because  large Zeeman energies are often required~\cite{maier2014majorana, luethi2023planar}. Only a few 
proposals 
have eliminated 
the need for  Zeeman terms, 
such as time-reversal invariant setups 
with Kramers pairs of MBSs~\cite{zhang2013time, nakosai2013majorana, keselman2013inducing, klinovaja2014time, haim2014time, gaidamauskas2014majorana, dumitrescu2014magnetic, haim2016interaction, schrade2017low, thakurathi2018majorana, aligia2018entangled}.
However, they complicate braiding, thus, systems with broken time-reversal symmetry are preferable.
For instance, a $\pi$-phase difference across superconductor-normal-superconductor (SNS) Josephson junctions requires only a reduced Zeeman energy to produce MBSs~\cite{pientka2017topological, hell2017two, setiawan2019topological, scharf2019tuning, dmytruk2019majorana, luethi2023planar} and enhanced orbital effects in, e.g., topological insulator nanowires enable MBSs without any Zeeman effect~\cite{cook2011majorana,vaitiekenas2020flux,legg2021majorana}. 
Interestingly, utilizing only phase differences in  planar SNSNS Josephson junctions (see Fig.~\ref{fig:setup}) it was recently shown that MBSs can exist in electron systems with linear SOI~\cite{lesser2022one}.
However, 
a significant mismatch in velocities of the inner and outer spin-split Fermi surfaces is required, which is difficult to achieve using  linear  SOI~\cite{lesser2022one}.

Here, instead we focus on holes in valence bands described by the Luttinger-Kohn Hamiltonian. We  
show analytically and numerically that achieving topological superconductivity in SNSNS Josephson junctions with CSOI requires also only 
phase differences, 
thereby extending the mechanism proposed for electrons~\cite{lesser2022one} to a different class of systems.
Moreover,  for such hole systems one finds conditions on the ideal junction geometry that enable  large topological regions of phase space, even for reduced junction transparencies.
Finally, an in-plane magnetic field provides  additional fingerprints of the topological phase. 
Using realistic parameters for Ge 2DHGs,
we argue that recent advances in superconductor-Ge devices enable MBSs to be realized in an experimentally accessible regime. 
Our results provide a roadmap to achieve topological superconductivity using only weak magnetic fields in a material with ultralong MFPs and large SOI. 

The structure of this paper is as follows. We introduce our setup as well as the corresponding Hamiltonian and show results for toy model parameters in Sec.~\ref{sec:snsns_with_CSOI}. In Sec.\ref{sec:realistic_parameters}, we present results for realistic parameters for Ge, and, furthermore, discuss the effect of an external Zeeman field. We conclude in Sec.~\ref{sec:conclusion}. In the Appendix, we give more information on the numerical calculations and give derivations of equations shown in the main text.

\section{\label{sec:snsns_with_CSOI}SNSNS Josephson junctions with CSOI}
\begin{figure*}
	\centering
	\includegraphics[width=\textwidth]{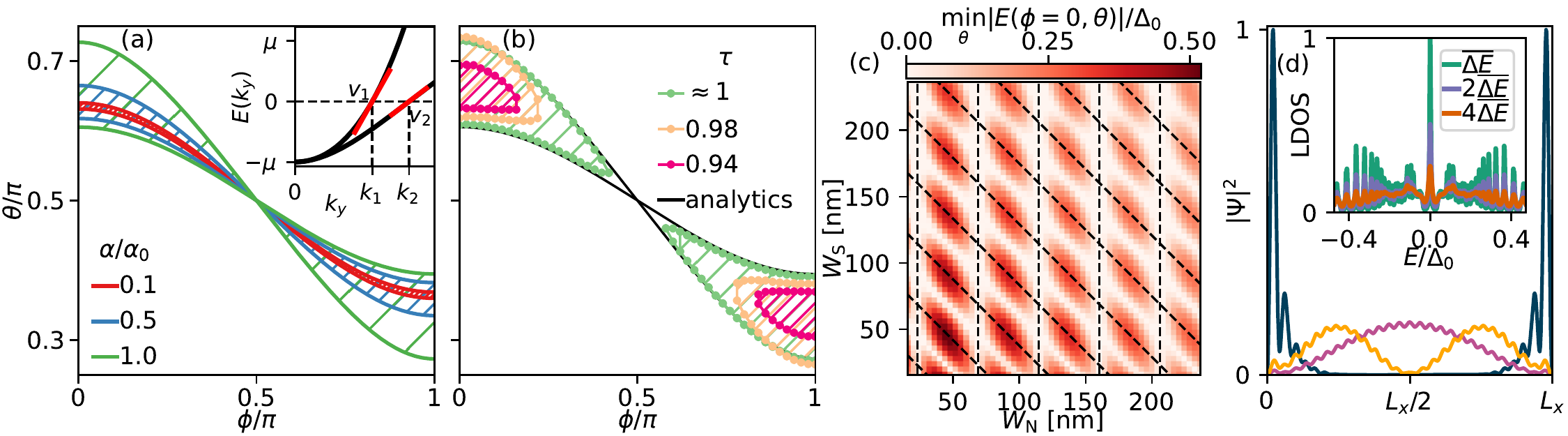}
	\caption{
		(a) Analytically calculated phase diagram
		[see Eq.~\eqref{eq:phase_diagram_analytical}] for different CSOI strengths~$\alpha$, compared to
		$\alpha_0=1600$~meV~nm$^3$~\cite{luethi2023planar},
		which is also the CSOI strength used in panels (b)-(d). 
		The topological phase is indicated by the hatched region.
		The two different Fermi velocities $v_1$ and $v_2$, and Fermi momenta $k_1$ and $k_2$ are shown in the inset.
		The larger  $\alpha$ is, the more $v_1$ and $v_2$ differ and therefore the larger is the topological phase.
		(b) Numerically calculated bulk gap closing points,  
		i.e., the points where $E(k_x=0)=0$, for different transparencies $\tau$ controlled via the tunneling barrier height $\mu_{\mathrm{b}}$. The topological phase is indicated by the hatched region. An imperfect transparency reduces the size of the topological phase region, therefore highly transparent junctions are favorable.
		We set $\mu_{\mathrm{b}}=0$ for $\tau\approx 1$,
		 $\mu_{\mathrm{b}}=1.3$~meV for $\tau=0.98$, and
		  $\mu_{\mathrm{b}}=6.0$~meV for $\tau=0.94$.
		(c) Minimum energy $\min_{\theta} |E(k_x=0, \phi=0, \theta)|$ in the semi-infinite geometry at $k_x=0$ and $\phi=0$. The tunneling barrier is fixed to
		 $2$~meV. 
		 The dashed lines indicate the ideal junction geometry conditions of Eq.~\eqref{eq:sweet_spot_condition}, for which a topological phase exists even
		 at a low transparency.
		 (d) Profile of the probability distribution $|\Psi|^2$ (arbitrary units) of the lowest energy state, i.e., the MBS (dark blue, $E/\Delta_{\mathrm{m}} = 7.4 \cdot 10^{-6}$), the first excited state (yellow, $E/\Delta_{\mathrm{m}} = 0.035$), and the third excited state (pink, $E/\Delta_{\mathrm{m}} = 0.037$), going through the left normal section in the finite geometry. 
		 The second excited state is not shown because it looks similar to the first excited state.
		 In contrast to all other calculations shown,  the junction here has a finite length $L_x=2$~$\mu$m in $x$ direction.
		 The inset shows the local density of states (LDOS) integrated over a small section at the junction end  for three different energy broadening coefficients ($\overline{\Delta E}$, $2\overline{\Delta E}$ and $4\overline{\Delta E}$, where $\overline{\Delta E}$ is the average level spacing), see Appendix \ref{appsec:ldos}. 
		 Although there are several in-gap states, the MBSs are well discernible as zero-energy peaks because the Andreev bound states are delocalized over the full length of the normal section.
		The parameters, taken from Ref.~\cite{luethi2023planar},  are $\hbar^2/2m^\ast = 580$~meV~nm$^2$,  
		$\alpha_a=0$, $\beta=4600$~meV~nm$^4$, $\mu=2.4$~meV, $\Delta_0=0.26$~meV in panels (a)-(c), and $\Delta_0=0.8$~meV in panel (d).
		The junction size in panels (a) and (b) is $W_{\mathrm{N}}=28$~nm and $W_{\mathrm{S}}=130$~nm. In panel (d) $W_{\mathrm{N}}=20$~nm and $W_{\mathrm{S}}=68$~nm.
		These parameters reveal the qualitative behavior of the system and are easily tractable numerically. We study realistic parameters later; see Fig.~\ref{fig:phase_diagram_realistic}. We give all parameters rounded to two significant digits. 
		All parameter values used for the numerical calculations are given in Appendix~\ref{app:parameters}.
		}
	\label{fig:toy_model_diagrams}
\end{figure*}

 Materials with CSOI have large differences in velocities  at the inner and outer Fermi surfaces; see Fig.~\ref{fig:toy_model_diagrams}(a). In particular, a 2DHG in Ge confined to the crystallographic $xy$ plane~\cite{winkler2003spin,marcellina2017spin} has 
 large and 
 dominant  CSOI of Rashba type
 with negligible linear SOI~\footnote{Although it is possible for systems with only linear (Rashba) SOI to have different Fermi velocities, e.g. due to  momentum dependence of the mass, the low filling of semiconductors means that such effects are typically weak. }.
 Anticipating superconductivity, 
 we introduce the Nambu basis $\Psi(x,y) = \Big(
	\psi_\uparrow(x,y) \, \,
	\psi_\downarrow(x,y) \, \,
	\psi_\uparrow^\dagger(x,y) \,\,
	\psi_\downarrow^\dagger(x,y)
\Big)^T$, where $\psi_s^\dagger(x,y)$ creates a particle at position $(x,y)$ with spin $s$. In this basis, the effective Hamiltonian of a Ge 2DHG, derived using the Luttinger-Kohn formalism, is~\cite{marcellina2017spin, terrazos2021theory, michal2021longitudinal}
\begin{align}
	&\!\mathcal{H}_{\mathrm{eff}} \!=\!\!
	\left[\!
	- \frac{\hbar^2}{2m^\ast} \left(\partial_x^2 + \partial_y^2\right)
	- \mu 
	\!\right] \tau_z
	+ 2i\alpha \Big[
	\partial_y \left(\partial_y^2-3\partial_x^2\right) \sigma_x 
	 \nonumber \\ &  
	 \!+ \partial_x \left(\partial_x^2-3\partial_y^2\right) \sigma_y \tau_z
	\Big]
	\! - \! 2i\alpha_a \left(\partial_x^2+\partial_y^2\right)\left(\partial_y \sigma_x + \partial_x \sigma_y \tau_z \right),
	\label{eq:effective_hamiltonian}
\end{align}
where $m^\ast$ is the effective mass, 
$\sigma_i$ ($\tau_i$) are the Pauli matrices acting in spin (particle-hole) space,
and $\alpha$ and $\alpha_a$ are the strengths of the CSOI. The term $\alpha_a$ comes from anisotropic corrections and for Ge $\alpha_a \ll \alpha$~\cite{marcellina2017spin}. As discussed in Ref.~\cite{luethi2023planar}, throughout we also introduce a quartic term $\mathcal{H}_4 = \beta (\partial_x^4 + \partial_y^4 + 2\partial_x^2\partial_y^2) \tau_z$ to avoid spurious additional Fermi surfaces in discretized numerical calculations. The exact size of $\beta$ does not affect our results. 
The energy spectrum of the full normal state Hamiltonian $\mathcal{H}=\mathcal{H}_{\mathrm{eff}} + \mathcal{H}_4$ in momentum space has two spin-split Fermi surfaces, labeled by $j=1$ ($j=2$) for the inner (outer) Fermi surface. The 
corresponding Fermi velocities 
are different; see  inset of Fig.~\ref{fig:toy_model_diagrams}(a).
These  velocities may be tuned by varying  the SOI strength or the chemical potential $\mu$.

\begin{figure*}
	\centering
	\includegraphics[width=\textwidth]{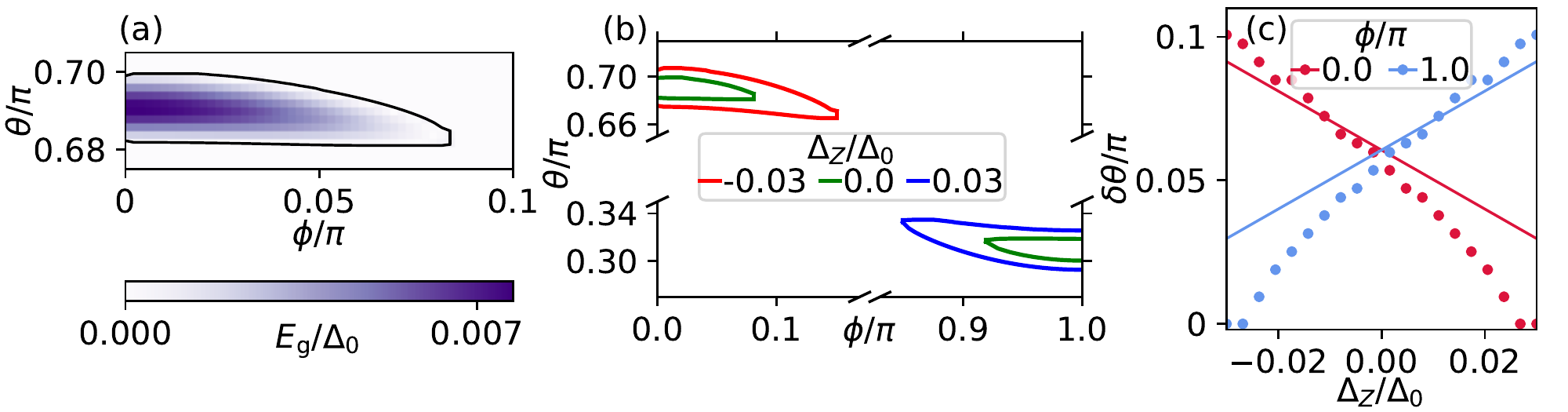}
	\caption{
		(a) Phase diagram for realistic Ge parameters, without Zeeman field. The black line indicates the topological phase transition. In the topological phase (purple area), we calculate the topological gap $E_{\mathrm{g}}$, which is defined in the semi-infinite geometry as $E_{\mathrm{g}} = \min_{k_x} |E(k_x)|$. For realistic parameters we find that there exists a finite topological region of phase space.
		(b) Phase diagram for realistic Ge parameters, with a small external Zeeman field $\Delta_Z$ in $x$ direction. 
		Without any Zeeman field, the phase diagram is symmetric under a $\pi$-rotation about $(\pi/2,\pi/2)$. Adding a Zeeman term breaks this symmetry, but increases the total area of the topological phase space.
		(c) The distance $\delta \theta$ between the two phase transition curves at different values of $\phi$ as a function of the Zeeman field $\Delta_Z$. The dots (lines) represent numerical (analytical) results. 
		Panels~(b) and~(c) demonstrate that the magnetic field breaks the rotational symmetry of phase space, giving an additional tool to distinguish MBSs from trivial states.
		For Ge, we use the following parameters~\cite{luethi2023planar, dantas2022determination}: $\hbar^2/2m^\ast=620$~meV~nm$^2$, $\alpha=190$~meV~nm$^3$, $\alpha_a=23$~meV~nm$^3$, $\mu= 7.4$~meV,
		$W_{\mathrm{S}}=170$~nm, $W_{\mathrm{N,l}}=W_{\mathrm{N,r}}=72$~nm, $\beta=74$~meV~nm$^4$, and $\mu_{\mathrm{b}}=12$~meV such that the transparency of the junction is $\tau=0.96$ (see Appendix~\ref{app:transparency} for more information about the transparency calculation). 
		We set the induced superconducting gap in germanium to $\Delta_{\mathrm{m}}=0.49$~meV, which is from Ref.~\cite{aggarwal2021enhancement}, where germanium is proximitized by superconducting aluminum and niobium.
		All parameters are rounded to two significant digits. 
		All parameter values used for the numerical calculations are given in Appendix~\ref{app:parameters}.
	}
	\label{fig:phase_diagram_realistic}
\end{figure*}

The SNSNS junction comprises three sections that are proximitized by superconductors and two normal sections between them; see Fig.~\ref{fig:setup}. 
The Hamiltonian for the proximity-induced superconducting potential is
\begin{equation}
	\mathcal{H}_{\mathrm{SC}} = i\sigma_y \left[
	 \Delta^\ast(y) \tau_+
	-  \Delta(y) \tau_-
	\right]/2
	\label{eq:sc_hamiltonian}
\end{equation}
with $\tau_\pm = \tau_x \pm i \tau_y$ and 
\begin{equation}
	\Delta(y) = \begin{cases}
		\Delta_{\mathrm{l}} e^{-i\theta} & \text{if } y < -W_{\mathrm{N,l}}-\frac{W_{\mathrm{S}}}{2}, \\
		\Delta_{\mathrm{m}}e^{i\phi} & \text{if } -\frac{W_{\mathrm{S}}}{2} \leq y < \frac{W_{\mathrm{S}}}{2}, \\
		\Delta_{\mathrm{r}}e^{i\theta} & \text{if } W_{\mathrm{N,r}}+\frac{W_{\mathrm{S}}}{2} \leq y, \\
		0 & \text{otherwise,}
	\end{cases}
	\end{equation}
where $W_{\mathrm{N,l}}$ ($W_{\mathrm{N,r}}$) is the width of the left (right) normal section, and $W_{\mathrm{S}}$ the width of the middle superconductor.

While a Zeeman field is not required to enter the topological phase, we will demonstrate later that it can be beneficial and provide additional features to distinguish MBSs from trivial bound states. The Hamiltonian for a magnetic field of strength $B$ applied in $x$ direction along the junction is $\mathcal{H}_Z = \Delta_Z(y) \sigma_x \tau_z$, where for simplicity we use
\begin{equation}
	\label{eq:zeeman_hamiltonian}
	\Delta_Z(y) = \begin{cases}
		\Delta_Z & \text{if } -W_{\mathrm{N,l}}-\frac{W_{\mathrm{S}}}{2} \leq y < -\frac{W_{\mathrm{S}}}{2}, \\
		& \text{or } \frac{W_{\mathrm{S}}}{2} \leq y < W_{\mathrm{N,r}} + \frac{W_{\mathrm{S}}}{2}, \\
		0 & \mathrm{otherwise,}
	\end{cases} 
\end{equation}
and $\Delta_Z = g \mu_B B$ with $g$ the $g$ factor of the material and $\mu_B$ the Bohr magneton. The induced superconducting gap is reduced as a magnetic field is applied, however, we focus on small magnetic fields and therefore neglect the reduction of the induced gap.
To take into account that the junction does not have perfect transparency, we introduce a potential barrier $	\mathcal{H}_{\mathrm{b}} = \mu_{\mathrm{b}}(y) \tau_z$, where
\begin{equation}
	\mu_{\mathrm{b}}(y) = \begin{cases}
		\mu_{\mathrm{b}} & \text{if } y_{\mathrm{b}}-\frac{W_{\mathrm{b}}}{2} \leq  y < y_{\mathrm{b}}+\frac{W_{\mathrm{b}}}{2} ,\\
		0 & \text{otherwise},\label{eq:tunneling_barrier}
	\end{cases}
\end{equation}
where $y_{\mathrm{b}} \in \{-W_{\mathrm{N,l}}-\frac{W_{\mathrm{S}}}{2}, -\frac{W_{\mathrm{S}}}{2}, 
	 \frac{W_{\mathrm{S}}}{2}, W_{\mathrm{N,r}}+\frac{W_{\mathrm{S}}}{2}\}$ and $W_{\mathrm{b}}$ is the width of the barrier.
The full Hamiltonian of the system is
\begin{equation}
	\mathcal{H} = 
	\mathcal{H}_{\mathrm{eff}} 
	+ \mathcal{H}_4
	+ \mathcal{H}_{\mathrm{SC}}
	+ \mathcal{H}_Z + \mathcal{H}_{\mathrm{b}}.
	\label{eq:full_hamiltonian}
\end{equation}

For now, neglecting the potential barrier and, for simplicity, assuming  that $W_{\mathrm{N,l}}=W_{\mathrm{N,r}}\equiv W_{\mathrm{N}}$ and that the Fermi velocities in both normal conducting sections are equal, the phase transition curves are given by
\begin{equation}
	\cos \left(\!\theta + \frac{2\Delta_Z W_{\mathrm{N}}(-1)^j}{v_j^{\mathrm{N}}} \!\right)
	+ \tanh \left(\frac{W_{\mathrm{S}} \Delta_{\mathrm{m}}}{v_j^{\mathrm{S,m}}}\right) \cos \left(\phi \right) = 0,
	\label{eq:phase_diagram_analytical}
\end{equation}
where $j=1$ ($j=2$) indicates the inner (outer) Fermi surface and $v_j^{\mathrm{N}}$ ($v_j^{\mathrm{S,m}}$) are the Fermi velocities of the corresponding Fermi surface in the normal conducting (middle superconducting) section. The topological phase is between the two curves defined by $j=1$ and $j=2$. 
In Appendix~\ref{app:zeeman_derivation}, we consider a more general case of $W_{\mathrm{N,l}} \neq W_{\mathrm{N,r}}$ as well as of different Fermi velocities in each section. 
 A larger CSOI strength~$\alpha$ 
results in a larger difference of  velocities at the Fermi level, which increases the topological region of phase space; see Fig.~\ref{fig:toy_model_diagrams}(a).

 To study the system numerically, the full Hamiltonian $\mathcal{H}$
is discretized. 
We utilize two different geometries: In the finite geometry, a finite extent in both $x$ and $y$ direction is assumed. In the semi-infinite geometry, it is assumed that the junction has a finite extent along the $y$ direction, but is infinitely extended along the $x$ direction, in this case, the momentum $k_x$ along the $x$ axis is a good quantum number. The discretized Hamiltonians for both cases are given in Appendix~\ref{app:discretized_ham}. In the following, for simplicity, we also assume a constant pairing potential, $\Delta_{\mathrm{l}}=\Delta_{\mathrm{m}}=\Delta_{\mathrm{r}}=\Delta_0$, in the superconducting sections.

The topological phase transition is of class~D~\cite{lesser2022one}, which is characterized by a closing of the bulk gap at momentum $k_x=0$~\cite{pientka2017topological}.
This allows us to calculate the phase diagram numerically; see Fig.~\ref{fig:toy_model_diagrams}(b). Imperfect transparency has a noticeable effect on the phase diagram, reducing the topological region of phase space. 
To estimate the transparency, we calculate the current-phase relation, see 
Appendix~\ref{app:transparency}. 
An imperfect transparency is caused by normal reflection at the SN interfaces~\cite{pientka2017topological}, which is neglected in the analytical derivation. However, assuming $W_{\mathrm{N,l}}=W_{\mathrm{N,r}} \equiv W_{\mathrm{N}}$ and Fermi momenta $k_1$ ($k_2$) for the inner (outer) Fermi surfaces within the junction, see inset of Fig.~\ref{fig:toy_model_diagrams}(a), values for $W_{\mathrm{N}}$ and $W_{\mathrm{S}}$ exist for which the effects of scattering are minimized (see Appendix~\ref{app:sweetspot}):
\begin{equation}
	W_{\mathrm{N}} = \frac{(2n+1)\pi}{k_1+k_2}, \,
	W_{\mathrm{S}}+W_{\mathrm{N}} = \frac{2m\pi}{k_1+k_2}, \,
	n,m \in \mathbb{Z}.
	\label{eq:sweet_spot_condition}
\end{equation}
We will refer to the conditions in Eq.~\eqref{eq:sweet_spot_condition} as the ideal junction geometry.
At $k_x=0$ and $\phi=0$ -- the region 
where the topological phase disappears last with decreasing transparency -- the minimum energy for varying $\theta$, $\mathrm{min}_\theta|E(k_x=0, \phi=0,\theta)|$, determines whether a topological phase can still exist at a certain value of the tunneling barrier $\mu_{\mathrm{b}}$. If the minimum is zero, a topological phase still exists. Therefore, at sufficiently high tunneling barriers, only systems with $W_{\mathrm{S}}$ and $W_{\mathrm{N}}$ close to the ideal junction geometry can still be topological; see Fig.~\ref{fig:toy_model_diagrams}(c). 
In an experiment, however, $W_{\mathrm{S}}$ and $W_{\mathrm{N}}$ are fixed after fabrication. In this case, the chemical potential can be tuned, which changes the Fermi momenta and therefore brings the system into a favorable 
configuration, where either one of the ideal junction geometry conditions is fulfilled, see Appendix~\ref{app:sweetspot}.
Systems in which both conditions of the ideal junction geometry are fulfilled are particularly favorable, as they allow for topological phases with particularly low transparencies. The deviation between the numerical and analytical ideal junction geometry in Fig.~\ref{fig:toy_model_diagrams}(c) is further discussed in Appendix~\ref{app:sweetspot}.

Although the topological gap of an SNSNS junction might not be large, as is also the case in SNS junctions~\cite{pientka2017topological, luethi2023planar}, MBSs are still clearly discernible as zero-energy peaks in the 
LDOS
at the junction ends because the low-energy Andreev bound states in planar Josephson junctions are delocalized over the full length of the normal section; see Fig.~\ref{fig:toy_model_diagrams}(d) and Appendix~\ref{appsec:mbs_wf}.

\section{\label{sec:realistic_parameters} Realistic parameters for germanium}
\raggedbottom
So far, using toy model parameters, we have demonstrated that an SNSNS junction based on a material with CSOI is a good candidate to host MBSs and we have shown the qualitative behavior of such systems. 
It was shown in Ref.~\cite{tosato2023hard} that fabricating Ge SNS junctions with transparencies up to 0.96 is possible. 
Using realistic parameters, we find a finite topological region of phase space, as shown in Fig.~\ref{fig:phase_diagram_realistic}(a).
We note that the topological gap in this realistic case is rather small, however, we discussed in Fig.~ \ref{fig:toy_model_diagrams}(d) that the first few excited states of planar Josephson junctions are spatially more extended than the MBSs. Therefore, a clear zero-energy peak in the local density of states at the junction ends is to be expected.

\subsection{\label{subsec:zeeman}Including a Zeeman field}
Given the small $g$ factor in Ge, a considerable advantage of a planar SNSNS junction compared to SNS junctions is that no Zeeman field is required to enter the topological phase. However, applying a magnetic field parallel to the junction does not destroy the topological region of phase space. In fact, a parallel magnetic field increases the total area of the topological phase region, see Fig.~\ref{fig:phase_diagram_realistic}(b) and Appendix~\ref{app:area_with_zeeman}. A magnetic field further breaks 
inversion symmetry which gives an additional tool to distinguish MBSs from trivial states; see Fig.~\ref{fig:phase_diagram_realistic}(c). Here, we plot the distance $\delta \theta$ between the two phase transition curves at a fixed $\phi$. Although the total topological area of phase space increases with a magnetic field, depending on $\phi$ and the sign of $\Delta_Z$, the difference $\delta \theta$ may increase or decrease. This is important because the antisymmetric behavior of the topological phase in a magnetic field provides an extra signature by which we can distinguish topological from trivial features. 
For instance, a zero-bias peak with a topological origin would be expected to be less robust against, e.g., changes in the superconducting phase difference $\theta$ for the magnetic field in one direction compared to the opposite direction, because $\delta \theta(\Delta_Z) \neq \delta \theta (-\Delta_Z)$.

\section{\label{sec:conclusion}Conclusion}
We demonstrated that the dominant cubic SOI in a Ge 2DHG
is ideal for hosting MBSs in SNSNS junctions. 
The SNSNS junction enters the topological phase without any Zeeman term, which eliminates the large magnetic fields that enable many trivial effects to mimic MBSs and provides a route to topological superconductivity in Ge despite the small $g$ factor.
Further, we show that an imperfect transparency is detrimental for the topological phase. However, we derive conditions on the ideal junction geometry for which a topological phase exists for experimentally achievable transparencies and parameters. The ideal junction geometry can be approached, e.g., by gating the junction.
Finally, although not necessary for MBSs within our setup, we also show that a Zeeman field increases the topological phase region  and provides an additional tool to distinguish MBSs from trivial states.

\begin{acknowledgments}
We thank Omri Lesser and Yuval Oreg for useful conversations. This work was supported by the Swiss National Science Foundation and NCCR SPIN (Grant No. 51NF40-180604). This project received funding from the European Union’s Horizon 2020 research and innovation program (ERC Starting Grant, Grant Agreement No. 757725).  H.F.L acknowledges support by the Georg H. Endress Foundation. K. L. acknowledges support by the Laboratory for Physical Sciences through the Condensed Matter Theory Center.
\end{acknowledgments}

\appendix

\section{\label{app:zeeman_derivation}Derivation of phase diagram with a Zeeman field}
In this Appendix, the analytical expression for the topological phase transition curve [see Eq.~\eqref{eq:phase_diagram_analytical}] is derived. 
The derivation generalizes the one of Ref.~\cite{lesser2022one} by including also a magnetic field in $x$ direction. Perfect transparency of the junction is assumed. Later,  in Appendix~\ref{app:sweetspot}, we will consider the impact of imperfect transparency and demonstrate the existence of an ideal junction geometry.

\begin{figure}
	\centering
	\includegraphics[width=0.5\textwidth]{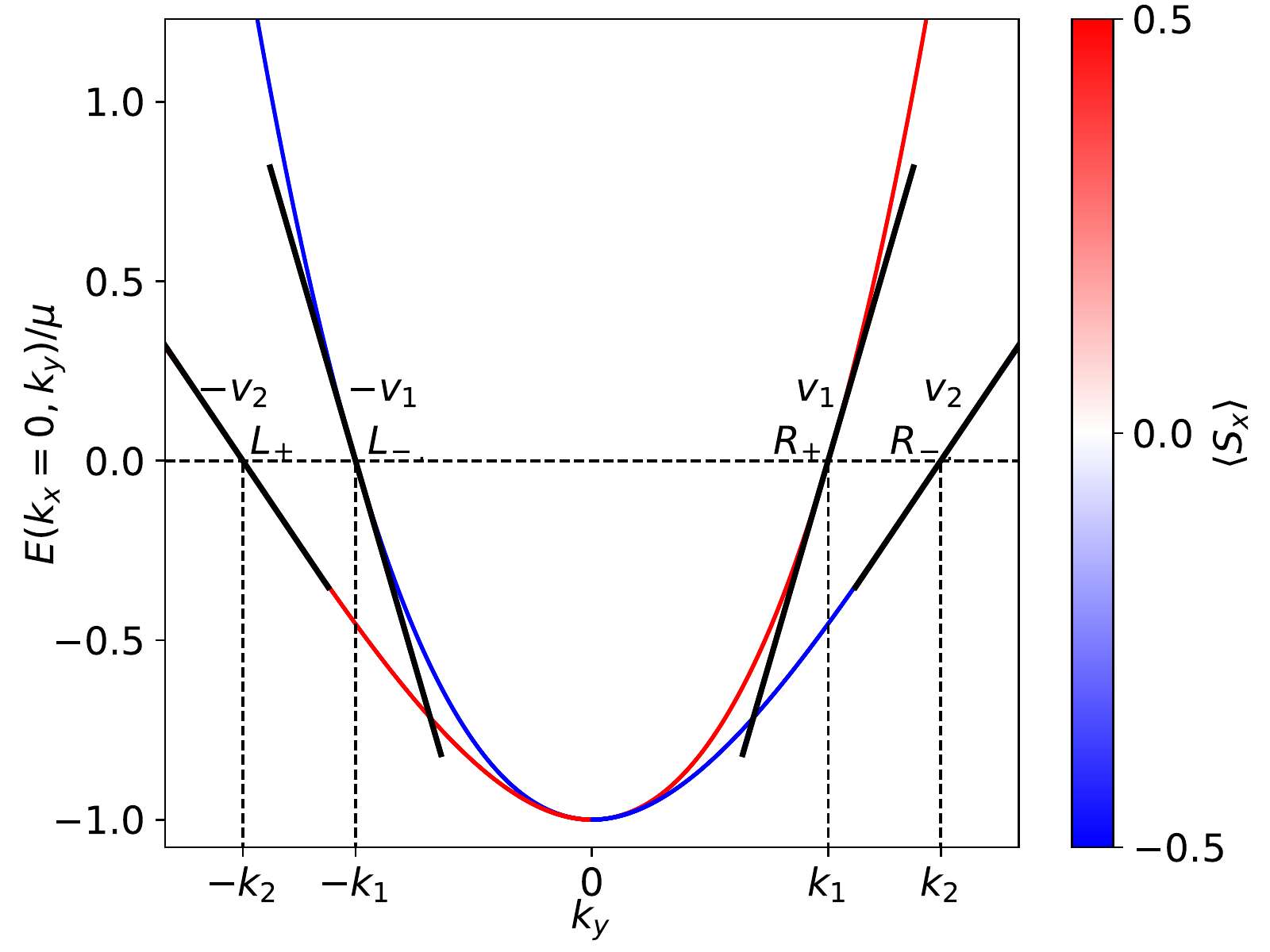}
	\caption{Spectrum of the effective Hamiltonian $\mathcal{H}_{\mathrm{eff}}+\mathcal{H}_4$ defined in the main text for $k_x=0$ with the linearized functions $L_\pm$ (left movers) and $R_\pm$ (right movers), the Fermi momenta $\pm k_{1,2}$, and the Fermi velocities $\pm v_{1,2}$. The color indicates the expectation value $\langle S_x \rangle$ of the spin in $x$ direction.
	}
	\label{fig:spectrum_linearization}
\end{figure}

Since a topological phase transition is characterized by a bulk gap closing at $k_x=0$, we set $k_x=0$ in the momentum space version of Eq.~\eqref{eq:effective_hamiltonian}
and linearize the spectrum, giving four branches; see Fig.~\ref{fig:spectrum_linearization}. We assume that the Zeeman term is only a perturbation to the spectrum and therefore linearize without the Zeeman term.
For $k_x=0$, the Hamiltonian commutes with $\sigma_x$ and therefore the spin along the $x$ direction $S_x$ is a good quantum number. Based on the direction of the Fermi velocity and the spin of the branch, we label the slow-varying fermionic fields as $L_\pm(k_y)$ and $R_\pm(k_y)$ ($L$ refers to left-movers, $R$ to right-movers, and the $\pm$ labels the spin eigenvalues) and their corresponding Fermi momenta and Fermi velocities as $k_j$ and $v_j$, $j=1,2$, respectively. The linearized spectrum is thus given by~\cite{laubscher2021majorana} 
\begin{align}
	H_\mathrm{kin} =&
	-v_2 L_+^\dagger \hat{k}_y L_+
	-v_1 L_-^\dagger \hat{k}_y L_-  
	\nonumber \\ &
	+v_1 R_+^\dagger \hat{k}_y R_+
	+v_2 R_-^\dagger \hat{k}_y R_-,
\end{align}
where $\hat{k}_y$ is the momentum operator in $y$ direction and we assume $v_j > 0$ and $k_j > 0$.
Since $\sigma_x$ commutes with the Hamiltonian, we will consider only states $\psi_\pm$ that are simultaneous eigenstates of $\sigma_x$ and the Hamiltonian, the index $\pm$ indicating their eigenvalue with respect to $\sigma_x$. Using the linearized ansatz, $\psi_\pm$ can be written as~\cite{klinovaja2012composite, laubscher2021majorana} 
\begin{align}
	\label{eq:psi_plus_minus_linearization}
	\psi_+(y) =& e^{-ik_2 y} L_+(y) + e^{i k_1 y} R_+(y) , \nonumber \\ 
	\psi_- (y)=& e^{-ik_1 y} L_- (y)+ e^{i k_2 y} R_- (y) .
\end{align}

To get the linearized version of the superconducting and the Zeeman term, we use
\begin{equation}
	\psi_{\uparrow} = \frac{1}{\sqrt{2}} \left( \psi_+ + \psi_- \right) ,  \quad
	\psi_{\downarrow} = \frac{1}{\sqrt{2}} \left( \psi_+ - \psi_- \right) ,
\end{equation}
where $\psi_{\uparrow/\downarrow}$ are spin eigenstates of $\sigma_z$. Assuming a  superconducting term with  pairing potential $\Delta$ and phase $\gamma$, the superconducting part becomes
\begin{align}
	H_{\mathrm{SC}} =& \frac{\Delta}{2} \Big\{
	\left[
	\psi_\uparrow \psi_\downarrow - \psi_\downarrow \psi_\uparrow
	\right] e^{i \gamma}  
	+ \left[
	\psi_\downarrow^\dagger \psi_\uparrow^\dagger
	- \psi_\uparrow^\dagger \psi_\downarrow^\dagger
	\right] e^{-i \gamma}
	\Big\} \nonumber \\
	=& \frac{\Delta}{2} \Big\{
	\left[
	L_- R_+
	- R_+ L_-
	+ R_- L_+  
	- L_+ R_-  
	\right] e^{i \gamma} 
	\nonumber \\ &
	+ \left[
	R_+^\dagger  L_-^\dagger 
	- L_-^\dagger R_+^\dagger
	+ L_+^\dagger R_-^\dagger 
	- R_-^\dagger  L_+^\dagger 
	\right] e^{-i\gamma}
	\Big\} \nonumber \\
	& + \text{ oscillating terms}\, ,
\end{align}
while the Zeeman term with Zeeman field $\Delta_Z$ reads 
\begin{align}
	H_Z =& \Delta_Z \left[
	\psi_\uparrow^\dagger \psi_\downarrow 
	+ \psi_\downarrow^\dagger \psi_\uparrow 
	\right] \nonumber \\
	=& \Delta_Z \left[
	L_+^\dagger L_+ 
	+ R_+^\dagger R_+
	- L_-^\dagger L_-
	- R_-^\dagger R_-
	\right] \nonumber \\ &
	+ \text{ oscillating terms}.
\end{align}

In the following, we neglect the fast oscillating terms~\cite{laubscher2021majorana}. In the basis $c(k_y) = \begin{pmatrix}
	R_+,& L_-^\dagger, &
	L_- ,& R_+^\dagger ,&
	R_- ,& L_+^\dagger, &
	L_+ ,& R_-^\dagger
\end{pmatrix}^T$,
the full Hamiltonian $H=H_{\mathrm{kin}}+H_{\mathrm{SC}}+H_Z$ is block-diagonal:
\begin{align}
	& H =
	\begin{pmatrix}
		\mathcal{H}(v_1 k_y) & 0 & 0 & 0 \\
		0 & -\mathcal{H}(v_1 k_y)  & 0 & 0 \\
		0 & 0 & -\mathcal{H}(-v_2 k_y) & 0 \\
		0 & 0 & 0 & \mathcal{H}(-v_2 k_y)
	\end{pmatrix}
	,
	\label{eq:full_linearized_ham} \\
	&\mathcal{H}(vk) = \begin{pmatrix}
		v k + \Delta_Z & \Delta e^{-i \gamma} \\
		\Delta e^{i\gamma} & -v k + \Delta_Z
	\end{pmatrix}.
	\label{eq:general_ham_block}
\end{align}

Since the Hamiltonian is block-diagonal, one can calculate the wave function for each block independently. Depending on the block, the Fermi velocity is either positive or negative. Thus, we assume a general Hamiltonian $\mathcal{H}(v k)$ as defined in Eq.~\eqref{eq:general_ham_block}, where the Fermi velocity $v=\pm v_1, \pm v_2$ 
can both be positive or negative.

First, consider the superconducting region:
We assume that there is no Zeeman term in this region, therefore giving the Hamiltonian
\begin{equation}
	\mathcal{H}(vk) = \begin{pmatrix}
		v k  & \Delta e^{-i \gamma} \\
		\Delta e^{i\gamma} & -v k
	\end{pmatrix}.
\end{equation}
As such, the energy in this region is given by
\begin{equation}
	E = \pm \sqrt{k^2 v^2 + \Delta^2} .
\end{equation}
Since we are interested in a closing of the bulk gap we set $E=0$, giving 
\begin{equation}
	k = \pm i \frac{\Delta}{v}
	\label{eq:k_in_sc}
\end{equation}
The corresponding eigenvector is
\begin{equation}
	\omega_\pm^{\mathrm{S}} = \begin{pmatrix}
		\pm i e^{-i \gamma} \\ 1
	\end{pmatrix}.
\end{equation}

In the normal section, we assume a Zeeman energy $\Delta_Z$ and thus the Hamiltonian is
\begin{equation}
	\mathcal{H}(v k) = \begin{pmatrix}
		v k + \Delta_Z & 0 \\
		0 & -v k + \Delta_Z
	\end{pmatrix}.
\end{equation}
The energy and eigenfunctions in this region are:
\begin{equation}
	E = \pm v k + \Delta_Z \stackrel{!}{=} 0
	\Rightarrow k = \pm \frac{\Delta_Z}{v} 
	\label{eq:k_in_n}
\end{equation}
and the eigenvectors are
\begin{equation}
	\omega_+^{\mathrm{N}} = \begin{pmatrix} 0 \\ 1 \end{pmatrix} , \quad
	\omega_-^{\mathrm{N}} = \begin{pmatrix} 1 \\ 0 \end{pmatrix}.
\end{equation}

We now make an ansatz for the wave function in each section of the SNSNS junction separately. We assume that the junction has perfect transparency, therefore there is no normal scattering. Thus, each block of the Hamiltonian defined in Eq.~\eqref{eq:full_linearized_ham} can be considered separately.
Generally, the chemical potential may be different in each section. Thus, we assume different Fermi momenta and velocities in each section, i.e., $k^{\mathrm{S,l}}$ and $v^{\mathrm{S,l}}$, $k^{\mathrm{S,m}}$ and $v^{\mathrm{S,m}}$, and $k^{\mathrm{S,r}}$ and $v^{\mathrm{S,r}}$ in the left, middle, and right superconductors, respectively, and $k^{\mathrm{N,l}}$ and $v^{\mathrm{N,l}}$ ($k^{\mathrm{N,r}}$ and $v^{\mathrm{N,r}}$) in the left (right) normal section. The widths of the normal sections are $W_{\mathrm{N,l}}$ and $W_{\mathrm{N,r}}$, respectively. The left and right superconducting sections are assumed to be infinitely extended, the middle superconducting section has a width $W_{\mathrm{S}}$. The induced superconducting gaps (superconducting phases) are $\Delta_{\mathrm{l}}$ ($-\theta$) in the left, $\Delta_{\mathrm{m}}$ ($\phi$) in the center, and $\Delta_{\mathrm{r}}$ ($\theta$) in the right superconducting sections; see Fig.~\ref{fig:setup}. 
We choose the coordinate system such that the interface between the left superconductor and the left normal section is at $y=0$. Note that this convention is different from the convention used in Fig.~\ref{fig:setup}, where $y=0$ is in the center of the middle superconductor.

Because the left superconductor is infinitely extended, only wave functions that are decaying for $y \rightarrow -\infty$ are valid. Therefore, the wave function in the left superconducting region contains either only $\omega_+^{S}$ or only $\omega_-^{S}$, depending on the sign of $v^{\mathrm{S,l}}$:
\begin{equation}
	\Psi_{\mathrm{S,l}}(y) = 
	e^{y \Delta_{\mathrm{l}}/|v^{\mathrm{S,l}}|} \begin{pmatrix} -i \, {\rm sgn}(v^{\mathrm{S,l}}) e^{i\theta} \\ 1 \end{pmatrix}. \label{wv}
\end{equation}
We emphasize that one gets the same type of the wave function for each of the four blocks defined by the Hamiltonian of Eq.~\eqref{eq:full_linearized_ham} separately. However, the basis in which Eq.~\eqref{wv} is defined depends on the block under consideration.
Following the same arguments, the wave function in the right superconducting section must be exponentially decaying for $y \rightarrow \infty$ and is therefore given by:
\begin{equation}
	\Psi_{\mathrm{S,r}}(y) = A_{\mathrm{S,r}}
	e^{-y \Delta_{\mathrm{r}}/|v^{\mathrm{S,r}}|} \begin{pmatrix} i \, {\rm sgn}(v^{\mathrm{S,r}}) e^{-i\theta} \\ 1 \end{pmatrix}.
\end{equation}
In the middle superconducting region, the ansatz for the wave function is:
\begin{align}
	\Psi_{\mathrm{S,m}}(y) =& 
	A_{\mathrm{S,m}} 
	e^{-y \Delta_{\mathrm{m}} /v^{\mathrm{S,m}}} \begin{pmatrix} i e^{-i\phi} \\ 1 \end{pmatrix} 
	\nonumber \\ &
	+ B_{\mathrm{S,m}} 
	e^{y \Delta_{\mathrm{m}} /v^{\mathrm{S,m}}} \begin{pmatrix} -i e^{-i\phi} \\ 1 \end{pmatrix}.
\end{align}
The ansatz for the wave function $\Psi_{\mathrm{N,l}}$ ($\Psi_{\mathrm{N,r}}$) in the left (right) normal region is
\begin{align}
	\Psi_{\mathrm{N,l/r}} =& 
	A_{\mathrm{N,l/r}} 
	e^{i y \Delta_Z/v^{N,l/r}} \begin{pmatrix} 0 \\ 1 \end{pmatrix}
	\nonumber \\ & 
	+ B_{\mathrm{N,l/r}} 
	e^{-i y \Delta_Z/v^{N,l/r}} \begin{pmatrix} 1 \\ 0 \end{pmatrix}.
\end{align}	
Next, all wave function parameters are determined by matching boundary conditions. As mentioned before, we assume perfect transparency and therefore the matching process can be done for each block defined by the Hamiltonian of Eq.~\eqref{eq:full_linearized_ham} separately as there is no back-scattering in this approach. Furthermore, since we consider a linearized spectrum, matching only the wave functions is sufficient, it is not required to match their first derivative.  At the first interface we get:
\begin{widetext}
	\begin{align}
		&\Psi_{\mathrm{S,l}}(0) = \Psi_{\mathrm{N,l}}(0) 
		\Leftrightarrow
		\begin{pmatrix} -i \, {\rm sgn}(v^{\mathrm{S,l}}) e^{i \theta} \\ 1 \end{pmatrix}
		= T_1 \begin{pmatrix} A_{\mathrm{N,l}} \\ B_{\mathrm{N,l}} \end{pmatrix} 
		\Leftrightarrow
		\begin{pmatrix} A_{\mathrm{N,l}} \\ B_{\mathrm{N,l}} \end{pmatrix} = T_1 \begin{pmatrix} -i \, {\rm sgn}(v^{\mathrm{S,l}}) e^{i \theta} \\ 1 \end{pmatrix} ,
		\label{eq:first_intersection} \\
		& T_1 = \begin{pmatrix} 0 & 1 \\ 1 & 0 \end{pmatrix},
	\end{align}
	using $T_1^{-1}=T_1$.
	
	The second interface is at $y=W_{\mathrm{N,l}}$, between the left normal section and the middle superconducting region. We require:
	\begin{align}
		&\Psi_{\mathrm{N,l}}(W_{\mathrm{N,l}}) = \Psi_{\mathrm{S,m}}(W_{\mathrm{N,l}})
		\Leftrightarrow
		T_1 D_N(W_{\mathrm{N,l}}, v^{\mathrm{N,l}}) 
		\begin{pmatrix} A_{\mathrm{N,l}} \\ B_{\mathrm{N,l}} \end{pmatrix}
		= 
		T_2 D_{S} (W_{\mathrm{N,l}}, \Delta_{\mathrm{m}}, v^{\mathrm{S,m}})
		\begin{pmatrix} A_{\mathrm{S,m}} \\ B_{\mathrm{S,m}} \end{pmatrix} \nonumber \\
		&\Leftrightarrow 
		\begin{pmatrix} A_{\mathrm{S,m}} \\ B_{\mathrm{S,m}} \end{pmatrix} = 
		D_{S} (-W_{\mathrm{N,l}}, \Delta_{\mathrm{m}}, v^{\mathrm{S,m}}) T_2^{-1} T_1 D_N(W_{\mathrm{N,l}}, v^{\mathrm{N,l}}) \begin{pmatrix} A_{\mathrm{N,l}} \\ B_{\mathrm{N,l}} \end{pmatrix},
		\label{eq:second_intersection} 
	\end{align}
	where we define
	\begin{align}
		T_2 =& \begin{pmatrix} ie^{-i\phi} & -ie^{-i\phi} \\ 1 & 1 \end{pmatrix}, \\
		D_{N}(W, v) =& \begin{pmatrix}
			e^{i W \Delta_Z/v} & 0 \\
			0 & e^{-i W \Delta_Z/v}
		\end{pmatrix}, \\
		D_{S}(W, \Delta, v) =& \begin{pmatrix}
			e^{-W \Delta/v} & 0 \\
			0 & e^{W \Delta/v}
		\end{pmatrix} ,
	\end{align}
	and using $D_{N}(W, v)^{-1} = D_N(-W,v)$ and $D_{S}(W, \Delta, v)^{-1}=D_{S}(-W, \Delta, v)$.
	
	The third intersection at $y=W_{\mathrm{N,l}}+W_{\mathrm{S}}$ is between the middle superconducting region and the right normal section:
	\begin{align}
		&\Psi_{\mathrm{S,m}}(W_{\mathrm{N,l}}+W_{\mathrm{S}}) = \Psi_{\mathrm{N,r}}(W_{\mathrm{N,l}}+W_{\mathrm{S}}) 
		\Leftrightarrow
		T_2 D_{S}(W_{\mathrm{N,l}}+W_{\mathrm{S}}, \Delta_{\mathrm{m}}, v^{\mathrm{S,m}})
		\begin{pmatrix} A_{\mathrm{S,m}} \\ B_{\mathrm{S,m}} \end{pmatrix}
		= 
		T_1 D_N(W_{\mathrm{N,l}}+W_{\mathrm{S}}, v^{\mathrm{N,r}})
		\begin{pmatrix} A_{\mathrm{N,r}} \\ B_{\mathrm{N,r}} \end{pmatrix} \nonumber \\
		&
		\Leftrightarrow \begin{pmatrix} A_{\mathrm{N,r}} \\ B_{\mathrm{N,r}} \end{pmatrix} =
		D_N(-W_{\mathrm{N,l}}-W_{\mathrm{S}},v^{\mathrm{N,r}})T_1 T_2 D_{S}(W_{\mathrm{N,l}}+W_{\mathrm{S}}, \Delta_{\mathrm{m}}, v^{\mathrm{S,m}})
		\begin{pmatrix} A_{\mathrm{S,m}} \\ B_{\mathrm{S,m}} \end{pmatrix}.
		\label{eq:third_intersection}
	\end{align}

	The fourth intersection is at $y=W_{\mathrm{N,l}}+W_{\mathrm{N,r}}+W_{\mathrm{S}}$ and is between the right normal section and the right superconducting region. Here, we require:
	\begin{align}
		& \Psi_{\mathrm{N,r}}(W_{\mathrm{N,l}}+W_{\mathrm{N,r}}+W_{\mathrm{S}}) = \Psi_{\mathrm{S,r}}(W_{\mathrm{N,l}}+W_{\mathrm{N,r}}+W_{\mathrm{S}}) \nonumber \\
		& \Leftrightarrow
		T_1 D_N(W_{\mathrm{N,l}}+W_{\mathrm{N,r}}+W_{\mathrm{S}}, v^{\mathrm{N,r}}) 
		\begin{pmatrix} A_{\mathrm{N,r}} \\ B_{\mathrm{N,r}} \end{pmatrix} 
		= 
		A_{\mathrm{S,r}} e^{-(W_{\mathrm{N,l}}+W_{\mathrm{N,r}}+W_{\mathrm{S}})\Delta/|v^{\mathrm{S,r}}|} \begin{pmatrix} i \, \mathrm{sgn}(v^{\mathrm{S,r}}) e^{-i\theta} \\ 1 \end{pmatrix} .
		\label{eq:matching_fourth}
	\end{align}
	
	Using Eqs.~\eqref{eq:first_intersection},~\eqref{eq:second_intersection},~\eqref{eq:third_intersection}, and $D_N(W_1, v) D_N(W_2, v) = D_N(W_1+W_2, v)$, $T_1 D_N(W,v) T_1 = D_N(-W,v)$, and equivalently for $D_{S}$, the left hand side of Eq.~\eqref{eq:matching_fourth} becomes:
	\begin{equation}
		D_N(-W_{\mathrm{N,r}}, v^{\mathrm{N,r}}) 
		T_2 
		D_{S}(W_{\mathrm{S}}, \Delta_{\mathrm{m}}, v^{\mathrm{S,m}})
		T_2^{-1} 
		D_N(-W_{\mathrm{N,l}}, v^{\mathrm{N,l}}) 
		\begin{pmatrix} -i \, \mathrm{sgn}(v^{\mathrm{S,l}}) e^{i \theta} \\ 1 \end{pmatrix}.
		\label{eq:matching_fourth_lhs}
	\end{equation}
	Now, we require that the ratio between the first and second element of the vector defined in Eq.~\eqref{eq:matching_fourth_lhs} equals the ratio on the right hand side of Eq.~\eqref{eq:matching_fourth}. This gives:
	\begin{equation}
		i \, {\rm sgn}(v^{\mathrm{S,r}}) e^{-i \theta} = -ie^{-2i W_{\mathrm{N,r}} \Delta_Z/v^{\mathrm{N,r}}} e^{-i\phi}\frac{
			e^{2iW_{\mathrm{N,l}} \Delta_Z/v^{\mathrm{N,l}}} \left(
			-1 + e^{2W_{\mathrm{S}} \Delta_{\mathrm{m}}/v^{\mathrm{S,m}}}
			\right)
			+ e^{i\theta} e^{i\phi} \mathrm{sgn}(v^{\mathrm{S,l}}) \left(
			1 + e^{2W_{\mathrm{S}} \Delta_{\mathrm{m}}/v^{\mathrm{S,m}}}
			\right)
		}{
			e^{2iW_{\mathrm{N,l}} \Delta_Z/v^{\mathrm{N,l}}} \left(
			1 + e^{2W_{\mathrm{S}} \Delta_{\mathrm{m}}/v^{\mathrm{S,m}}}
			\right)
			+ e^{i\theta} e^{i\phi} \mathrm{sgn}(v^{\mathrm{S,l}}) \left(
			-1 + e^{2W_{\mathrm{S}} \Delta_{\mathrm{m}}/v^{\mathrm{S,m}}}
			\right)
		}  .
	\end{equation}
	
	After some simplification, and using ${\rm sgn}(v^{\mathrm{N,l}})={\rm sgn}(v^{\mathrm{N,r}})={\rm sgn}(v^{\mathrm{S,l}})={\rm sgn}(v^{\mathrm{S,m}})={\rm sgn}(v^{\mathrm{S,r}})$, we obtain:
	\begin{equation}
		\label{eq:phase_diagram_general}
		\cos \left(\theta - \Delta_Z \left[\frac{W_{\mathrm{N,l}}}{v^{\mathrm{N,l}}}+\frac{W_{\mathrm{N,r}}}{v^{\mathrm{N,r}}} \right]\right)
		+ \tanh \left(\frac{W_{\mathrm{S}} \Delta_{\mathrm{m}}}{|v^{\mathrm{S,m}}|}\right) \cos \left(\phi - \Delta_Z \left[\frac{W_{\mathrm{N,l}}}{v^{\mathrm{N,l}}}-\frac{W_{\mathrm{N,r}}}{v^{\mathrm{N,r}}} \right] \right) = 0.
	\end{equation}
\end{widetext}
For simplicity, we assume $\Delta_{\mathrm{l}}=\Delta_{\mathrm{m}}=\Delta_{\mathrm{r}}=\Delta_0$ in all numerical calculations.
We assume $v^{\mathrm{N,l}}=v^{\mathrm{N,r}}$ and $v^{\mathrm{S,l}}=v^{\mathrm{S,r}}$. Then, to get the equation for the first branch, set $v^{\mathrm{N,l}}=v^{\mathrm{N,r}}\equiv v_1^\mathrm{N}$ and $v^{\mathrm{S,m}}=v_1^{\mathrm{S,m}}$ [see Eq.~\eqref{eq:full_linearized_ham}]. For the equation of the second branch, set $v^{\mathrm{N,l}}=v^{\mathrm{N,r}}=-v_2^{\mathrm{N}}$ and $v^{\mathrm{S,m}}=-v_2^{\mathrm{S,m}}$ [see Eq.~\eqref{eq:full_linearized_ham}]. This leads us to Eq.~\eqref{eq:phase_diagram_analytical}. 
To conclude, Eq.~ \eqref{eq:phase_diagram_general} gives two distinct phase transition curves. This is to be expected from the Hamiltonian defined in Eq.~\eqref{eq:full_linearized_ham}, as it has four blocks, but the first and second, as well as the third and fourth block, are particle-hole partners. Therefore, two independent solutions remain. We emphasize again that the topological phase is the area between the two phase transition curves, thus it is crucial that the two solutions have different Fermi velocities, i.e., $v_1 \neq v_2$.

\subsection{\label{app:area_with_zeeman}Area of topological phase with Zeeman field}
The area $A$ of the topological phase is given by:
\begin{equation}
	A = \int_0^{2\pi} d\phi \, \left|
	\theta_1 \left(\phi\right) - \theta_2 \left(\phi\right)
	\right|,
	\label{eq:definition_area_topo_phase}
\end{equation}
with $\theta_{1,2}(\phi)$ defined by Eq.~\eqref{eq:phase_diagram_analytical} 
(assuming $W_{\mathrm{N,l}}=W_{\mathrm{N,r}} \equiv W_{\mathrm{N}}$). To get an analytical estimate of the area, assume $\tanh\left(W_{\mathrm{S}} \Delta_{\mathrm{m}}/v_j^{\mathrm{S,m}}\right) \ll 1$, such that the 
$\arccos$-function can be expanded:
\begin{equation}
	\theta_j(\phi) \approx \frac{\pi}{2} + \cos \phi \tanh \left(\frac{W_{\mathrm{S}} \Delta_{\mathrm{m}}}{v_j^{\mathrm{S,m}}}\right) -  \frac{2 (-1)^j W_{\mathrm{N}}\Delta_Z}{v_j^{\mathrm{N}}}.
\end{equation}
Therefore, the integral in Eq.~\eqref{eq:definition_area_topo_phase} is of the form
\begin{align}
	A=& \left|b_1\right| \int_0^{2\pi} d \phi \left|\cos \phi + b_2 \right| 
	\nonumber \\ &
	= \left|b_1\right| \left[ 4\sqrt{1-b_2^2} + 2b_2\pi - 4b_2\arccos \left(b_2\right) \right] \, \text{if } |b_1|<1,
	\label{eq:area_with_zeeman_step_1}
\end{align}
with
\begin{align}
	b_1 =& 
	\tanh\left(\frac{W_{\mathrm{S}} \Delta_{\mathrm{m}}}{v_1^{\mathrm{S,m}}}\right)
	- \tanh\left(\frac{W_{\mathrm{S}} \Delta_{\mathrm{m}}}{v_2^{\mathrm{S,m}}}\right),\\
	b_2 =& -\frac{2 W_{\mathrm{N}} \Delta_Z \left(\frac{1}{v_1^{\mathrm{N}}}+\frac{1}{v_2^{\mathrm{N}}}\right)}{b_1}.
\end{align}
Assuming that $W_{\mathrm{N}} \Delta_Z/v_j^{\mathrm{N}} \ll 1$,
Eq.~\eqref{eq:area_with_zeeman_step_1} can be expanded in powers of $\Delta_Z$, giving
\begin{align}
	A \approx& 4\left|
	\tanh\left(\frac{W_{\mathrm{S}} \Delta_{\mathrm{m}} }{v_1^{\mathrm{S,m}}}\right)
	-\tanh\left(\frac{W_{\mathrm{S}} \Delta_{\mathrm{m}}}{v_2^{\mathrm{S,m}}}\right)
	\right|
	\nonumber \\ &
	+ \frac{8\left(\frac{1}{v_1^{\mathrm{N}}}+\frac{1}{v_2^{\mathrm{N}}}\right)^2 W_{\mathrm{N}}^2}{\left|\tanh\left(\frac{W_{\mathrm{S}} \Delta_{\mathrm{m}} }{v_1^{\mathrm{S,m}}}\right)
		-\tanh\left(\frac{W_{\mathrm{S}} \Delta_{\mathrm{m}} }{v_2^{\mathrm{S,m}}}\right)\right|} \Delta_Z^2,
\end{align}
which means that the area increases quadratically with $\Delta_Z$. 
However, it was assumed that the induced superconducting gaps were independent of $\Delta_Z$. Realistically, the induced gap decreases when a magnetic field is applied. Depending on the system parameters, this decrease of the superconducting gap can lead to an overall decrease of the area of the topological phase. However, since we consider only small magnetic fields (compared to the critical field of the superconductor), we assume that the decrease of the superconducting gap is negligible.

\section{\label{app:discretized_ham}Discretized Hamiltonian}

\subsection{\label{app:discretized_ham_finite}Finite geometry}
The Hamiltonians given 
in the main text are discretized on a square lattice with a lattice spacing~$a$. 
The Nambu basis in the finite geometry is given by
\begin{equation}
	\label{eq:nambu_finite}
	c_{n,m} =
	\begin{pmatrix}
		c_{\uparrow, n, m} & c_{\downarrow, n, m} & c_{\uparrow, n, m}^\dagger & c_{\downarrow, n, m}^\dagger
	\end{pmatrix}^T,
\end{equation}
where $n,m \in \mathbb{Z}$ and $c_{s,n,m}^\dagger$ creates a particle with spin $s$ at position $(x,y) = (na, ma)$. The full Hamiltonian in the finite geometry is~\cite{luethi2023planar}:
\begin{equation}
	\bar{H} = \frac{1}{2} \left(
	\bar{H}_\mathrm{eff} + \bar{H}_{\mathrm{b}} + \bar{H}_4 
	+ \bar{H}_{\mathrm{SC}} + \bar{H}_{Z} \right)
	\label{eq:tb_finite_geometry}
\end{equation}
\begin{align}
	&\bar{H}_\mathrm{eff} + \bar{H}_{\mathrm{b}} + \bar{H}_4  \nonumber \\ 
	&=
	\sum_{n=0}^{N_x-1} \sum_{m=0}^{N_y-1} \!\!
	c_{n, m}^\dagger \left(
	\frac{2t}{a^2} - \frac{\mu}{2}  + \frac{\mu_{\mathrm{b},m}}{2}
	+ \!\! \frac{10 \beta}{a^4}
	\right) \tau_z c_{n,m} 
	\nonumber \\ &
	+ \!\! \sum_{n=1}^{N_x-1} \sum_{m=0}^{N_y-1} \!\!
	c_{n, m}^\dagger \left[ -\frac{t}{a^2} - \frac{4i(\alpha+\alpha_a)}{a^3}\sigma_y 
	-\frac{8\beta}{a^4} \right] \tau_z  c_{n-1,m} \nonumber \\
	&+ \!\! \sum_{n=0}^{N_x-1} \sum_{m=1}^{N_y-1} \!\!
	c_{n, m}^\dagger \left[ -\frac{t}{a^2} \tau_z  - \frac{4i(\alpha+\alpha_a)}{a^3}\sigma_x 
	-\frac{8\beta }{a^4} \tau_z \right]  \! c_{n,m-1}
	\nonumber \\ &
	+ \!\! \sum_{n=2}^{N_x-1} \sum_{m=0}^{N_y-1} \!\!
	c_{n, m}^\dagger \left[\frac{i(-\alpha+\alpha_a)}{a^3}  \sigma_y 
	+ \frac{\beta}{a^4}\right] \tau_z c_{n-2,m} \nonumber \\
	&+ \!\! \sum_{n=0}^{N_x-1} \sum_{m=2}^{N_y-1} \!\!
	c_{n, m}^\dagger \left[ \frac{i(-\alpha+\alpha_a)}{a^3} \sigma_x 
	+ \frac{\beta}{a^4} \tau_z \right] c_{n,m-2}
	\nonumber \\  &
	+  \!\! \sum_{n=1}^{N_x-1} \sum_{m=1}^{N_y-1} \!\!
	c_{n, m}^\dagger \!\! \left[ \frac{i(3\alpha+\alpha_a)}{a^3}  \! \left(\sigma_x \!
	+ \! \sigma_y \tau_z \right) \! + \! \frac{2\beta}{a^4} \tau_z \right] \! c_{n-1,m-1}  \nonumber \\
	&+ \!\! \sum_{n=0}^{N_x-2} \sum_{m=1}^{N_y-1} \!\!
	c_{n, m}^\dagger \!\! \left[  \frac{i(3\alpha+\alpha_a)}{a^3} \!  \left(\sigma_x  \!
	-\! \sigma_y \tau_z \right) \! + \! \frac{2\beta}{a^4} \tau_z \right] \! c_{n+1,m-1}
	\nonumber \\  &
	+ \mathrm{H.c.} ,
	\label{eq:tb_effective_ham}
\end{align}
with $N_x$ and $N_y$ the number of lattice points in $x$ and $y$ direction respectively, $t=\hbar^2/2m^\ast$, and
\begin{equation}
	\mu_{\mathrm{b},m} = \begin{cases}
		\mu_{\mathrm{b}} & \text{if } m_{\mathrm{b}} - \frac{N_{\mathrm{b}}}{2} \leq m < m_{\mathrm{b}} + \frac{N_{\mathrm{b}}}{2}, \\
		0 & \text{otherwise,}
	\end{cases}
	\label{eq:definition_mu_b_discrete}
\end{equation}
where $m_{\mathrm{b}} \in \{
N_{\mathrm{S,l}}, 
N_{\mathrm{S,l}}+N_{\mathrm{N,l}},
N_{\mathrm{S,l}}+N_{\mathrm{S,m}}+N_{\mathrm{N,l}},
N_y-N_{\mathrm{S,r}}
\}$,
$N_{\mathrm{S,l}}$ ($N_{\mathrm{S,r}}$) is the number of lattice points in $y$ direction in the left (right) superconductor, $N_{\mathrm{S,m}}$ is the number of lattice points in $y$ direction in the middle superconductor, $N_{\mathrm{N,l}}$ ($N_{\mathrm{N,r}}$) is the number of lattice points in $y$ direction in the left (right) normal conducting regions,  $N_y=N_{\mathrm{S,l}}+N_{\mathrm{S,m}}+N_{\mathrm{S,r}}+N_{\mathrm{N,l}}+N_{\mathrm{N,r}}$, 		
and $N_{\mathrm{b}}$ is the number of lattice points in $y$ direction in the barrier. Throughout we set $N_{\mathrm{b}}=2$.
We relate the widths to the number of lattice points as follows:
\begin{align}
	L_x =& \left(N_x - 1\right) a, \\
	L_y =& \left(N_y - 1\right) a , \\
	W_{\mathrm{N,l}} =& \left(N_{\mathrm{N,l}}-1\right) a \label{eq:relation_WN_NN_left}, \\
	W_{\mathrm{N,r}} =& \left(N_{\mathrm{N,r}}-1\right) a \label{eq:relation_WN_NN_right} ,\\
	W_{\mathrm{S}} =& \left(N_{\mathrm{S,m}}-1\right) a \label{eq:relation_WS_NS} .
\end{align}

Furthermore, the term describing the induced pairing potential due to the superconductor is
\begin{align}
	\bar{H}_{\mathrm{SC}} &= \sum_{n=0}^{N_x-1} \sum_{m=0}^{N_y-1} 
	c_{n, m}^\dagger
	\frac{i \sigma_y}{2} 
	\left( 
	\Delta_m^\ast 
	\tau_+
	- \Delta_m 
	\tau_- 
	\right) 
	c_{n,m} ,
	\label{eq:tb_sc}
\end{align}
\begin{align}
	\Delta_m &= 
	\begin{cases}
		\Delta_{\mathrm{l}} e^{-i \theta} & \text{if } 0 \leq m < N_{\mathrm{S,l}}, \\
		\Delta_{\mathrm{m}} e^{i \phi} & \text{if } N_{\mathrm{S,l}}+N_{\mathrm{N,l}} \leq m < N_{\mathrm{S,l}}+N_{\mathrm{S,m}}+N_{\mathrm{N,l}}, \\
		\Delta_{\mathrm{r}} e^{i\theta} & \text{if } N_y-N_{\mathrm{S,r}} \leq m < N_y, \\
		0              & \text{otherwise}.
	\end{cases}
	\label{eq:delta_sc_definition} 
\end{align}
The Zeeman term is given by
\begin{align}
	\bar{H}_Z &= \sum_{n=0}^{N_x-1} \sum_{m=0}^{N_y-1} c_{n,m}^\dagger
	\Delta_{Z,m} \sigma_x \tau_z
	c_{n,m}, \\
	\Delta_{Z,m} &= \begin{cases}
		\Delta_Z & \text{if } N_{\mathrm{S,l}} \leq m < N_{\mathrm{S,l}}+N_{\mathrm{N,l}}, \\
		& \text{or } N_y-N_{\mathrm{S,r}}-N_{\mathrm{N,r}} \leq m < N_y-N_{\mathrm{S,r}},\\
		0              & \text{otherwise.}
	\end{cases} 
	\label{eq:delta_z_definition}
\end{align}
We note that the finite geometry is only used for Figs.~\ref{fig:toy_model_diagrams}(d) and~\ref{fig:wf_full_2d}. All other calculations are done in the semi-infinite geometry.

\subsection{\label{app:discretized_ham_semi_inf}Semi-infinite geometry}
For the semi-infinite geometry, we assume that the junction has an infinite extent in $x$ direction (with periodic boundary conditions). Therefore, the momentum $k_x$ along the $x$ axis is a good quantum number. The corresponding Nambu basis is:
\begin{equation}
	\label{eq:nambu_semi_inf}
	c_{k_x,m} = 
	\begin{pmatrix}
		c_{\uparrow, k_x,m} &
		c_{\downarrow, k_x,m} &
		c_{\uparrow, -k_x,m}^\dagger &
		c_{\downarrow, -k_x,m}^\dagger
	\end{pmatrix}^T,
\end{equation}
where $c_{s, k_x,m}^\dagger$ creates a particle at position $y=ma$ with spin $s$ and momentum $k_x$ in the $x$ direction. The Hamiltonian in the semi-infinite geometry is~\cite{luethi2023planar}:
\begin{widetext}
\begin{align}
	\tilde{H} =  \frac{1}{2} \int_{-\infty}^\infty \!\! dk_x
	\left[ 
	\tilde{H}_\mathrm{eff} \!\left(k_x\right) 
	+ \tilde{H}_{\mathrm{b}} \right.&\left. \left(k_x\right) 
	+ \tilde{H}_4 \left(k_x\right) 
	+ \tilde{H}_{SC}  \left(k_x\right) 
	+ \tilde{H}_{Z}  \left(k_x\right) 
	\right]  ,
	\label{eq:semi_inf_geometry_integral} \\
	\tilde{H}_\mathrm{eff}\left(k_x\right) 
	+ \tilde{H}_{\mathrm{b}}\left(k_x\right) 
	+\tilde{H}_4\left(k_x\right) =&
	\sum_{m=0}^{N_y-1}
	c_{k_x, m}^\dagger
	\left\{
	\frac{t\left[2-\cos\left(k_x a\right)\right]}{a^2}
	-\frac{\mu}{2} + \frac{\mu_{\mathrm{b},m}}{2}
	+\frac{\beta}{a^4} \left[10-8\cos\left(k_x a\right)+\cos\left(2k_x a\right)\right]
	\right. \nonumber \\ &  \left.
	-\frac{4\left(\alpha+\alpha_a\right)}{a^3} \sin\left(k_x a\right)\sigma_y + \frac{-\alpha+\alpha_a}{a^3} \sin\left(2k_x a\right) \sigma_y 
	\right\} \tau_z c_{k_x, m} \nonumber \\
	&
	+\sum_{m=1}^{N_y-1} c_{k_x,m}^\dagger \left\{
	-\frac{t}{a^2} \tau_z 
	+\frac{4\beta}{a^4} \left[-2+\cos\left(k_x a\right)\right]  \tau_z
	+ \frac{2 \left(3 \alpha+\alpha_a\right) }{a^3} \sin\left(k_x a\right) \sigma_y \tau_z 
	\right. \nonumber \\&  \left.
	+ \frac{2i }{a^3} \left[\left(3\alpha+\alpha_a\right) \cos\left(k_x a\right)-2\left(\alpha+\alpha_a\right)\right] \sigma_x 
	\right\} c_{k_x, m-1} \nonumber\\
	&+\sum_{m=2}^{N_y-1} c_{k_x,m}^\dagger
	\left(
	\frac{i (-\alpha+\alpha_a)}{a^3} \sigma_x
	+\frac{\beta}{a^4} \tau_z
	\right) c_{k_x, m-2}
	+ \mathrm{H.c.}.
	\label{eq:h_eff_semi_infinite}  
\end{align}
\end{widetext}
The induced superconducting pairing potential is described by
\begin{align}
	\tilde{H}_{SC}\left(k_x\right) &=  \sum_{m=0}^{N_y-1} 
	c_{k_x,m}^\dagger 
	\frac{i \sigma_y}{2} \left(
	\Delta_m \tau_+
	- \Delta_m^\ast \tau_-
	\right) 
	c_{k_x,m},
\end{align}
where 
$\Delta_m$ is defined in Eq.~\eqref{eq:delta_sc_definition}. The Zeeman term is:
\begin{equation}
	\tilde{H}_Z(k_x) = \sum_{m=0}^{N_y-1} 
	\Delta_{Z,m}
	c_{k_x,m}^\dagger 
	\sigma_x \tau_z
	c_{k_x,m},
\end{equation}
where $\Delta_{Z,m}$ is defined in Eq.~\eqref{eq:delta_z_definition}.

\subsection{\label{app:parameters}Parameters}
In this subsection, we give the numerical values for all parameters used to generate Figs.~\ref{fig:toy_model_diagrams} and~\ref{fig:phase_diagram_realistic}. 
For Fig.~\ref{fig:toy_model_diagrams}, all parameters are given in units of $t=\hbar^2/2m^\ast$ and the lattice spacing $a$. 
The parameters are $a=4$~nm, $\alpha=0.68 ta$, $\beta=0.49 ta^2$, $\mu=0.065 ta^{-2}$. In Fig.~\ref{fig:toy_model_diagrams}(a)-\ref{fig:toy_model_diagrams}(c) $\Delta_0 = 0.007 ta^{-2}$, $N_{\mathrm{N,l}}=N_{\mathrm{N,r}}=8$, and $N_{\mathrm{S}}=33$. In Fig.~\ref{fig:toy_model_diagrams}(d) $\Delta_0=0.022 ta^{-2}$, $N_{\mathrm{N,l}}=N_{\mathrm{N,r}}=6$, and $N_{\mathrm{S}}=18$. 
In Fig.~\ref{fig:toy_model_diagrams}(a), the outer superconductors are assumed to be infinitely wide (in $y$ direction), while in Figs.~\ref{fig:toy_model_diagrams}(b) and \ref{fig:toy_model_diagrams}(c), their widths are finite, given by $N_\textrm{S,l}=N_\textrm{S,r}=300$. In Fig.~\ref{fig:toy_model_diagrams}(d) the widths of the outer superconductors are $N_\textrm{S,l}=N_\textrm{S,r}=100$ and the system has a finite extent in $x$ direction of $N_x=500$.
The potential barriers in Fig.~\ref{fig:toy_model_diagrams}(b) are $\mu_{\textrm{b}}=0.035 ta^{-2}$ and $\mu_{\textrm{b}}=0.165 ta^{-2}$. The potential barrier in Fig.~\ref{fig:toy_model_diagrams}(c) is $\mu_{\textrm{b}}=0.055 ta^{-2}$ and in Fig.~\ref{fig:toy_model_diagrams}(d) $\mu_{\textrm{b}}=0$.
The superconducting coherence lengths $\xi_j = v_j/\Delta_0$ in Figs.~\ref{fig:toy_model_diagrams}(a)-\ref{fig:toy_model_diagrams}(c) are $\xi_1= 50 a$ and $\xi_2= 41a$. In \ref{fig:toy_model_diagrams}(d) $\xi_1=30a$ and $\xi_2= 13a$. For Figs.~\ref{fig:toy_model_diagrams}(a)-\ref{fig:toy_model_diagrams}(d), the Fermi wavelengths $\lambda_j=2\pi/k_j$ are $\lambda_1 = 28a$ and $\lambda_2= 19a$.

For Fig.~\ref{fig:phase_diagram_realistic} the parameters are instead given in units of the lattice spacing $a$ and the energy scale $E_0 = 37$~meV, which is an energy scale discussed in Ref.~\cite{luethi2023planar}. In these units, the parameters are $a=1.85$~nm, $t=\hbar^2/2m^\ast=4.9 E_0 a^2$,
$\alpha=0.81 E_0 a^3$, $\alpha_\mathrm{a}=0.1 E_0 a^3$, $\beta=0.17 E_0 a^4$, $\mu=0.2 E_0$, $\Delta_0 = 0.013 E_0$, $\mu_{\textrm{b}}=0.33 E_0$, $N_{\mathrm{N,l}}=N_{\mathrm{N,r}}=40$, $N_{\mathrm{S}}=95$, and $N_\mathrm{S,l}=N_\mathrm{S,r} = 500$.
The superconducting coherence lengths are $\xi_1=161a$ and $\lambda_2 = 143a$ and the Fermi wavelengths are $\lambda_1 = 32a$ and $\lambda_2 = 30a$.

\section{\label{app:transparency} Transparency calculation}
\begin{figure}
	\centering
	\includegraphics[width=\linewidth]{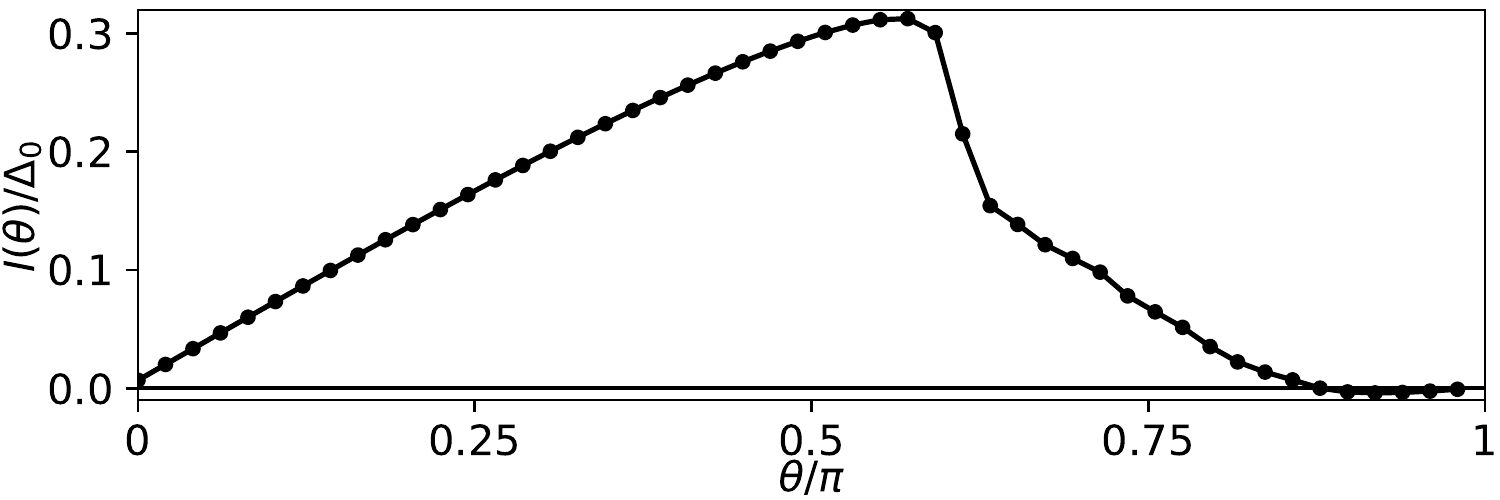}
	\caption{Current-phase relation of an SNSNS junction, where the superconducting phase differences are $-\theta/2$, $\phi=0$, $\theta/2$ for the left, middle, and right superconductors respectively. This current-phase relation is more complicated than the simple form for an SNS junction defined in Eq.~\eqref{eq:definition_transparency}.}
	\label{fig:current_phase_snsns}
\end{figure}

In this Appendix, we show how the transparency of the junction is calculated.
We estimate the transparency using the current-phase relation $I(\rho)$, which is proportional to~\cite{beenakker1991universal}:
\begin{equation}
	I(\rho) \propto \frac{\partial}{\partial \rho} \int_{E<0} d k_x  E(k_x, \rho)
	\label{eq:current_phase_numerics},
\end{equation}
where $\rho$ is the superconducting phase difference and $E(k_x, \rho)$ is the energy spectrum calculated in the semi-infinite geometry. 
The current-phase relation of an SNSNS junction is not uniquely defined because there are two independent superconducting phase differences. For simplicity, we set $\phi=0$. This, however, results in a rather complicated current-phase relation for an SNSNS junction; see Fig.~\ref{fig:current_phase_snsns}. It is not clear how this current-phase relation depends on the transparency. In contrast, for an SNS junction, the transparency is well approximated by~\cite{beenakker1991universal}:
\begin{equation}
	I(\rho) = I_0 \frac{\sin(\rho)}{\sqrt{1-\tau \sin^2(\rho/2)}},
	\label{eq:definition_transparency}
\end{equation}
where $I_0$ is a real parameter. Therefore, we estimate the transparency of the system using the current-phase relation of an SNS junction with normal conducting section width $W_{\mathrm{N}}\equiv W_{\mathrm{N,l}}=W_{\mathrm{N,r}}$. The transparency $\tau$ is then obtained by comparing the ratio of the first and second harmonics of Eqs.~\eqref{eq:current_phase_numerics} and~\eqref{eq:definition_transparency}.

\section{\label{app:sweetspot} Ideal junction geometry}

\begin{figure}
	\centering
	\includegraphics[width=0.8\linewidth]{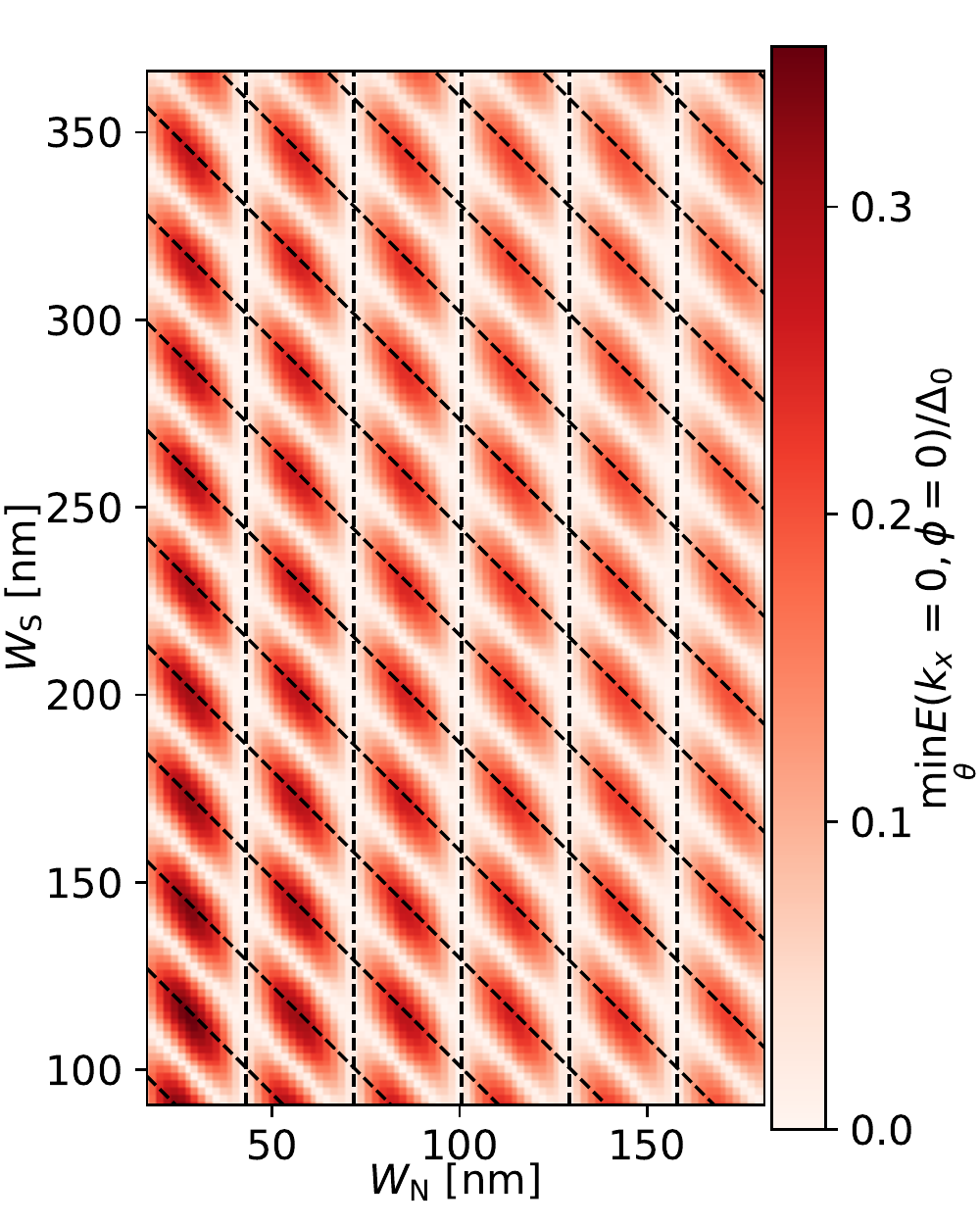}
	\caption{
		Minimum energy varying $\theta$ in the semi-infinite geometry at $k_x = 0$ and $\phi = 0$. This is the same plot as Fig.~\ref{fig:toy_model_diagrams}(c), however here using realistic parameters for Ge. The parameters are as in Fig.~\ref{fig:phase_diagram_realistic} and the tunneling barrier is fixed to $5.18$~meV. 
	}
	\label{fig:min_energy}
\end{figure}

The analytic calculation of the phase diagram (see Ref.~\cite{lesser2022one} and Appendix~\ref{app:zeeman_derivation}) assumes perfect transparency, which is apparent from the Hamiltonian in Eq.~\eqref{eq:full_linearized_ham}, as it does not couple left and right movers. Phenomenologically, such a coupling can be introduced by adapting the linearized Hamiltonian of Appendix~\ref{app:zeeman_derivation} as follows:
\begin{equation}
	H^\prime = H + c^\dagger (k_y)
	\begin{pmatrix}
		0 & 0 & 0 & H_T \\
		0 & 0  & H_T & 0 \\
		0 & H_T & 0 & 0 \\
		H_T & 0 & 0 & 0
	\end{pmatrix}
	c(k_y),
\end{equation}
where the Hamiltonian $H$ is defined in Eq.~\eqref{eq:full_linearized_ham} and 
\begin{equation}
	H_T = \begin{pmatrix}
		\delta & 0 \\
		0 & \delta
	\end{pmatrix},
\end{equation}
where $\delta$ labels the overlap between the left- and right-movers. In this ansatz, we assume that spin is conserved. Perturbatively, the overlap $\delta$ can be estimated as the overlap between the left- and right-moving wave functions calculated for the uncoupled system in Appendix~\ref{app:zeeman_derivation}. 
For the following derivation, we focus on the spin $+$ branch, noting that the derivation for the spin $-$ branch is equivalent. By defining the wave functions
\begin{align}
	\Psi_R =& \begin{pmatrix}
		R_+(k_y) & L_-^\dagger(-k_y) 
	\end{pmatrix}^T, \\
	\Psi_L =& \begin{pmatrix}
		L_+(k_y) & R_-^\dagger(-k_y)
	\end{pmatrix}^T,
\end{align}
we want a condition for which parameters the scattering has a negligible impact. This is the case if there is no overlap between left- and right-movers:
\begin{equation}
	0 = \langle \Psi_R | V | \Psi_L \rangle,
	\label{eq:initial_condition_sweet_spot}
\end{equation}
where $V$ is the potential that couples the left- and right-movers, i.e., the potential barrier at the SN interface. For simplicity, we assume a Dirac $\delta$ potential at each intersection:
\begin{align}
	V(y) =& V_0 \left[
	\delta(y)
	+ \delta(y-W_{\mathrm{N}})
	\right. \nonumber \\ & \left.
	+ \delta(y-W_{\mathrm{N}}-W_{\mathrm{S}})
	+ \delta(y-2W_{\mathrm{N}}-W_{\mathrm{S}})
	\right],
\end{align}
where $V_0$ is real, $W_{\mathrm{N}}=W_{\mathrm{N,l}}=W_{\mathrm{N,r}}$ is the width of each normal conducting section, and $W_{\mathrm{S}}$ is the width of the middle superconducting section. Since the largest extent of the topological phase is at $\phi=0$ (or $\phi=\pi$), we set $\phi=0$. Furthermore, we set $\Delta_Z=0$ and assume that the Fermi velocities and Fermi momenta are equal in all sectors, i.e., $v^{\mathrm{S,l}}_j=v^{\mathrm{S,m}}_j=v^{\mathrm{S,r}}_j=v^{\mathrm{N,l}}_j=v^{\mathrm{N,r}}_j \equiv v_j$ and $k^{\mathrm{S,l}}_j=k^{\mathrm{S,m}}_j=k^{\mathrm{S,r}}_j=k^{\mathrm{N,l}}_j=k^{\mathrm{N,r}}_j \equiv k_j$ for $j=1,2$. 
We note that in this step, the oscillating prefactors $e^{\pm i k_{1,2} y}$ in Eq.~\eqref{eq:psi_plus_minus_linearization} must be taken into account explicitly. 
At the first interface, which is at $y=0$, the wave functions are [see Eq.~\eqref{eq:first_intersection}]:
\begin{equation}
	\Psi_L (0) = \begin{pmatrix}
		i e^{i\theta_L} \\ 1
	\end{pmatrix} , \quad
	\Psi_R(0) = \begin{pmatrix}
		-i e^{i \theta_R} \\ 1
	\end{pmatrix},
\end{equation}
where $\theta_L$ and $\theta_R$ can be expressed using Eq.~\eqref{eq:phase_diagram_analytical}:
\begin{align}
	e^{i\theta_L} &= -\tanh\left(\frac{W_{\mathrm{S}} \Delta_{\mathrm{m}}}{v_2}\right) + i \sqrt{1-\tanh^2 \left(\frac{W_{\mathrm{S}} \Delta_{\mathrm{m}}}{v_2}\right)}, \\
	e^{-i\theta_R} &= -\tanh\left(\frac{W_{\mathrm{S}} \Delta_{\mathrm{m}}}{v_1}\right) - i \sqrt{1-\tanh^2 \left(\frac{W_{\mathrm{S}} \Delta_{\mathrm{m}}}{v_1}\right)}.
\end{align}
At the second interface at $y=W_{\mathrm{N}}$, one gets:
\begin{align}
	\Psi_L (W_{\mathrm{N}}) = e^{-ik_2 W_{\mathrm{N}}} \Psi_L(0), \\
	\Psi_R(W_{\mathrm{N}}) =e^{ik_1 W_{\mathrm{N}}} \Psi_R(0).
\end{align}
At the third interface at $y=W_{\mathrm{N}}+W_{\mathrm{S}}$ the wave functions are [see Eq.~\eqref{eq:third_intersection}]:
\begin{align}
	\Psi_L (W_{\mathrm{N}}+W_{\mathrm{S}}) = e^{-ik_2(W_{\mathrm{N}}+W_{\mathrm{S}})} M(-v_2) \Psi_L(0), \\
	\Psi_R(W_{\mathrm{N}}+W_{\mathrm{S}}) = e^{ik_1(W_{\mathrm{N}}+W_{\mathrm{S}})} M(v_1) \psi_R(0) ,
\end{align}
with
\begin{equation}
	M(v) = \begin{pmatrix}
		\cosh \left(\frac{W_{\mathrm{S}}\Delta_{\mathrm{m}}}{v}\right)
		& -i \sinh \left(\frac{W_{\mathrm{S}}\Delta_{\mathrm{m}}}{v}\right) \\
		i \sinh \left(\frac{W_{\mathrm{S}}\Delta_{\mathrm{m}}}{v}\right) &
		\cosh \left(\frac{W_{\mathrm{S}}\Delta_{\mathrm{m}}}{v}\right)
	\end{pmatrix}.
\end{equation}
Finally, at the fourth interface at $y=2W_{\mathrm{N}}+W_{\mathrm{S}}$, one obtains
\begin{align}
	\Psi_L(2W_{\mathrm{N}}+W_{\mathrm{S}}) = e^{-ik_2 W_{\mathrm{N}}} \Psi_L(W_{\mathrm{N}}+W_{\mathrm{S}}) , \\
	\Psi_R(2W_{\mathrm{N}}+W_{\mathrm{S}}) = e^{ik_1 W_{\mathrm{N}}} \Psi_R(W_{\mathrm{N}}+W_{\mathrm{S}}) .
\end{align}
Therefore, Eq.~\eqref{eq:initial_condition_sweet_spot} becomes:
\begin{align}
	0 = &
	\left(1+e^{-i(k_1+k_2)W_{\mathrm{N}}}\right)
	\begin{pmatrix}
		ie^{-i\theta_R} & 1
	\end{pmatrix}
	\nonumber \\
	& \times
	\left( 1 
	+ e^{-i(k_1+k_2)(W_{\mathrm{N}}+W_{\mathrm{S}})} [M(v_1)]^\dagger M(-v_2)
	\right)
	\begin{pmatrix}
		i e^{i\theta_L} \\ 1
	\end{pmatrix} .
\end{align}
Finally, this gives the condition that the overlap of left- and right-movers is minimized when 
\begin{align}
	0 = &
	2i e^{-i(k_1+k_2)(W_{\mathrm{N}}+W_{\mathrm{S}})}
	\frac{e^{W_{\mathrm{S}}\Delta_{\mathrm{m}}/v_1}-e^{W_{\mathrm{S}}\Delta_{\mathrm{m}}/v_2}}{\left(1+ie^{W_{\mathrm{S}}\Delta_{\mathrm{m}}/v_1}\right) \left(1-ie^{W_{\mathrm{S}}\Delta_{\mathrm{m}}/v_2}\right)}
	\nonumber \\ & \times
	\left(1+e^{-i(k_1+k_2)W_{\mathrm{N}}}\right)
	\left(-1+e^{i(k_1+k_2)(W_{\mathrm{N}}+W_{\mathrm{S}})}\right).
\end{align}
Thus, either one of the following conditions must be fulfilled:
\begin{equation}
	0 = 1+e^{-i(k_1+k_2)W_{\mathrm{N}}}\;\;{\rm or}\;\;\,
	0 = -1+e^{i(k_1+k_2)(W_{\mathrm{N}}+W_{\mathrm{S}})},
\end{equation}
which are the conditions in Eq.~\eqref{eq:sweet_spot_condition}. 
We show the existence of an ideal junction geometry for the toy-model parameters in Fig.~\ref{fig:toy_model_diagrams}(c) 
and for realistic Ge parameters in Fig.~\ref{fig:min_energy}. In both cases, the analytically calculated lines [see Eq.~\eqref{eq:sweet_spot_condition}] 
deviate systematically from the numerically calculated values. There are several reasons why the numerically observed ideal junction geometry is slightly different compared to the analytically expected conditions. For instance, although we define a relation between $N_{\mathrm{S}}$ and $W_{\mathrm{S}}$ (and between $N_{\mathrm{N}}$ and $W_{\mathrm{N}}$) in Eqs.~\eqref{eq:relation_WN_NN_left}-\eqref{eq:relation_WS_NS}, it is only a convention and not uniquely defined. A further complication comes from the potential barrier, which is not a Dirac delta potential, but has a finite extent in the numerical calculation. 

\begin{figure}[!t]
	\centering
	\includegraphics[width=\linewidth]{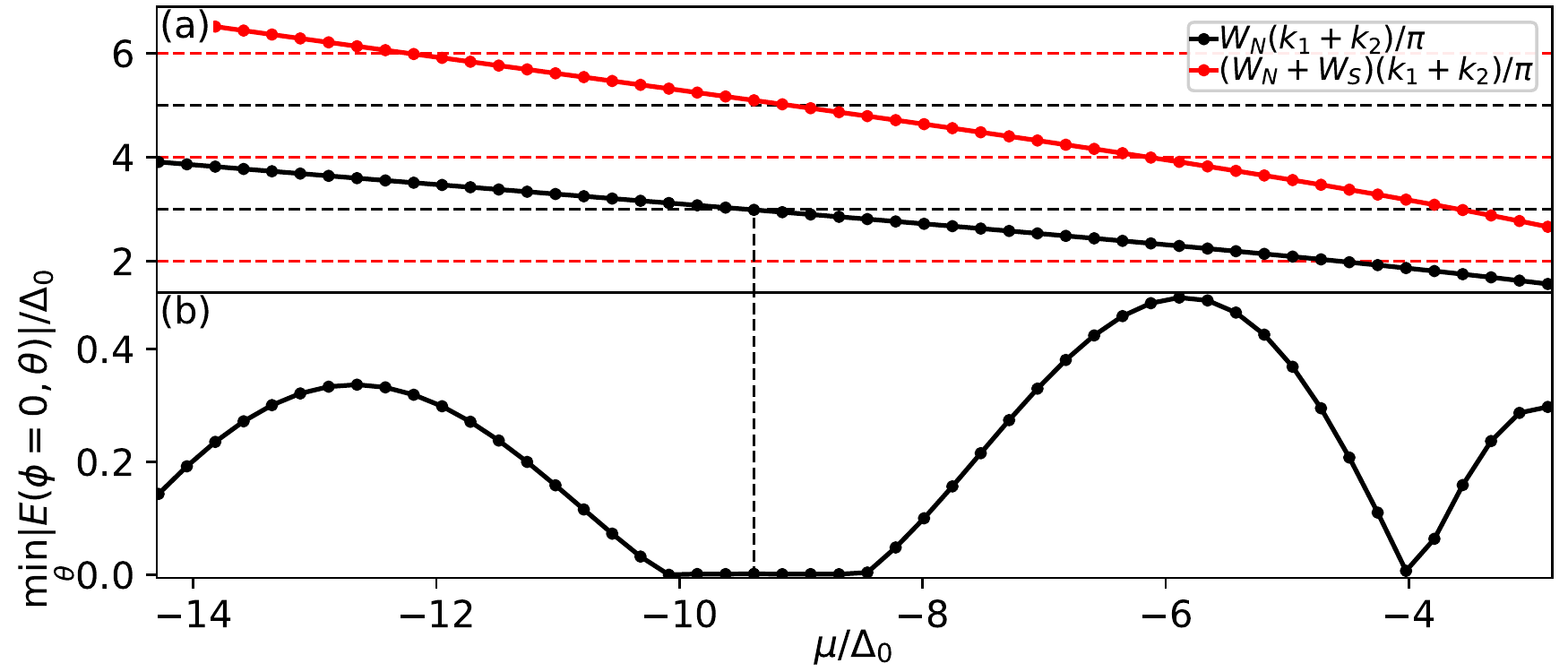}
	\caption{
		For fixed widths $W_\mathrm{N}$ and $W_\mathrm{S}$, the chemical potential $\mu$ is varied to tune the system such that one of the conditions of Eq.~\eqref{eq:sweet_spot_condition}
		is fulfilled.
		(a) The black dots indicate $\frac{W_\textrm{N}(k_1+k_2)}{\pi}$, which has to be an odd integer (indicated by the horizontal black dashed lines) to satisfy one of the ideal junction geometry conditions in Eq.~\eqref{eq:sweet_spot_condition}.
		The red dots indicate $\frac{(W_\textrm{N}+W_\textrm{S})(k_1+k_2)}{\pi}$, which has to be an even integer (indicated by the horizontal red dashed lines) to satisfy Eq.~\eqref{eq:sweet_spot_condition}.
		By varying the chemical potential $\mu$, the Fermi momenta $k_1$ and $k_2$ are varied, while $W_\textrm{N}$ and $W_\textrm{S}$ are kept constant.
		(b) Minimum energy $\min_\theta |E(\phi=0, \theta)|$ in the semi-infinite geometry at $k_x=0$ and $\phi=0$. A topological phase only exists if this minimum energy is zero.
		The vertical dashed line indicates where one of the conditions of Eq.~\eqref{eq:sweet_spot_condition} 
		is satisfied.
		The parameters are as in Fig.~\ref{fig:toy_model_diagrams}(c). The width of the normal section is $W_\mathrm{N}=68$~nm (i.e., $N_\mathrm{N}=13$) and $W_\mathrm{S}=48$~nm (i.e., $N_\mathrm{S}=18$).
	}
	\label{fig:sweet_spot_varying_mu}
\end{figure}

Although no longer giving simple analytical conditions, we have checked numerically that qualitatively, the same conditions on the ideal junction geometry still apply even for the case when $W_{\mathrm{N,l}} \neq W_{\mathrm{N,r}}$, or having different tunneling barrier heights at each intersection, or when there are different Fermi velocities and wave vectors in each section.

We mention in the main text that in an experiment, $W_\textrm{N}$ and $W_\textrm{S}$ are fixed after fabrication. However, by gating the sample, the chemical potential $\mu$ is changed, which affects the Fermi momenta $k_1$ and $k_2$. This changes the conditions on the ideal junction geometry, see Eq.~\eqref{eq:sweet_spot_condition}. 
While it is not always reasonable to change the chemical potential such that both conditions in Eq.~\eqref{eq:sweet_spot_condition}
are fulfilled, it is possible to change the chemical potential such that at least one of the conditions in Eq.~\eqref{eq:sweet_spot_condition} 
is satisfied, see Fig.~\ref{fig:sweet_spot_varying_mu}.

\section{\label{appsec:ldos}Local density of states}

\begin{figure}[!t]
	\centering
	\includegraphics[width=\linewidth]{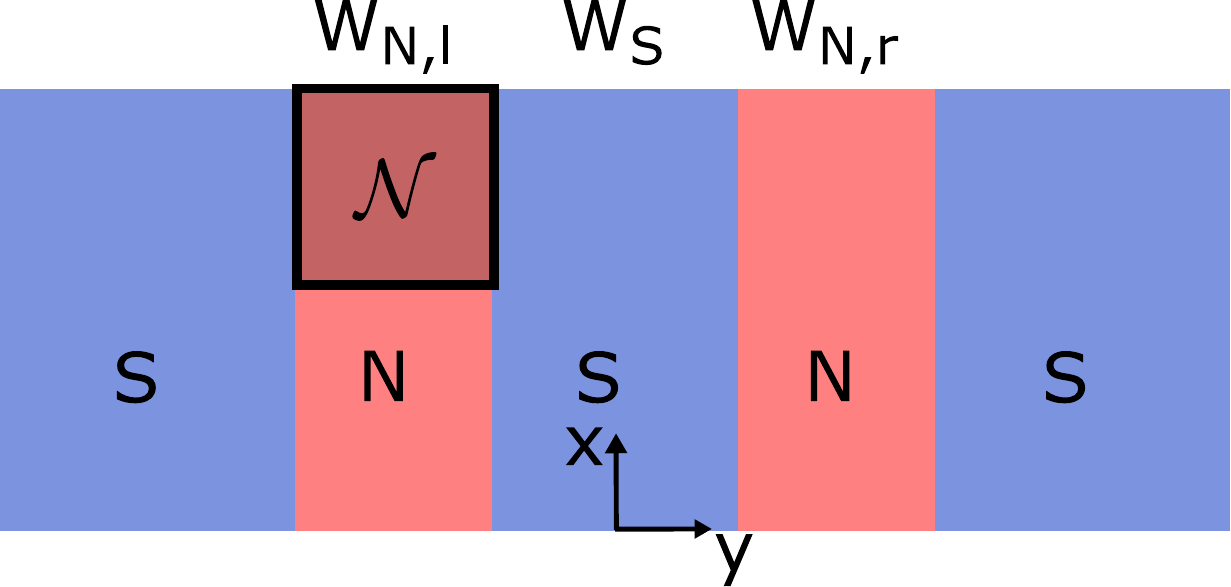}
	\caption{Definition of the area $\mathcal{N}$ in Eq.~\eqref{eq:ldos_definition}.   It is a square of side length $W_{\mathrm{N,l}}$. }
	\label{fig:definitions_s_ares}
\end{figure}

The integrated local density of states $\mathrm{LDOS}$ for Fig.~\ref{fig:toy_model_diagrams}(d) is defined as follows:
\begin{equation}
	\mathrm{LDOS} = \sum_n \iint_\mathcal{N} dx \, dy \int_{-\mathcal{E}}^{\mathcal{E}} dE \, f(E-E_n) \left|\Psi_n (x, y) \right|^2,
	\label{eq:ldos_definition}
\end{equation}
where $n$ labels all eigenstates of the system, their energy being $E_n$ and their wave function $\Psi_n$, $\mathcal{N}$ is the area over which to integrate, $\mathcal{E}$ is a parameter that defines the boundary of the energy integral [we use $\mathcal{E}/\Delta_0=0.47$ for Fig.~\ref{fig:toy_model_diagrams}(d)], and $f(E-E_n)$ is a broadening function. Using the same coordinate system as in the main text, we define 
$\mathcal{N}=\{(x,y) | \, 0\leq x < W_{\mathrm{N,l}} \text{ and } -W_{\mathrm{N,l}}-W_{\mathrm{S}}/2 \leq y < -W_{\mathrm{S}}/2  \}$, 
see Fig.~\ref{fig:definitions_s_ares}. The broadening function is defined as a Cauchy distribution:
\begin{equation}
	f(E) = \frac{1}{\pi} \frac{\nu}{E^2+\nu^2},
\end{equation}
where $\nu$ is the broadening coefficient. In the inset of Fig.~\ref{fig:toy_model_diagrams}(d), the broadening coefficient of the green curve is $\overline{\Delta E}$, $2\overline{\Delta E}$ for the blue curve, and $4 \overline{\Delta E}$ for the orange curve. Here, $\overline{\Delta E}$ is the numerically determined average level spacing, which is $1.9$~$\mu$eV.

\section{\label{appsec:mbs_wf}Wave function of Majorana bound states}

\begin{figure}[!t]
	\centering
	\includegraphics[width=\linewidth]{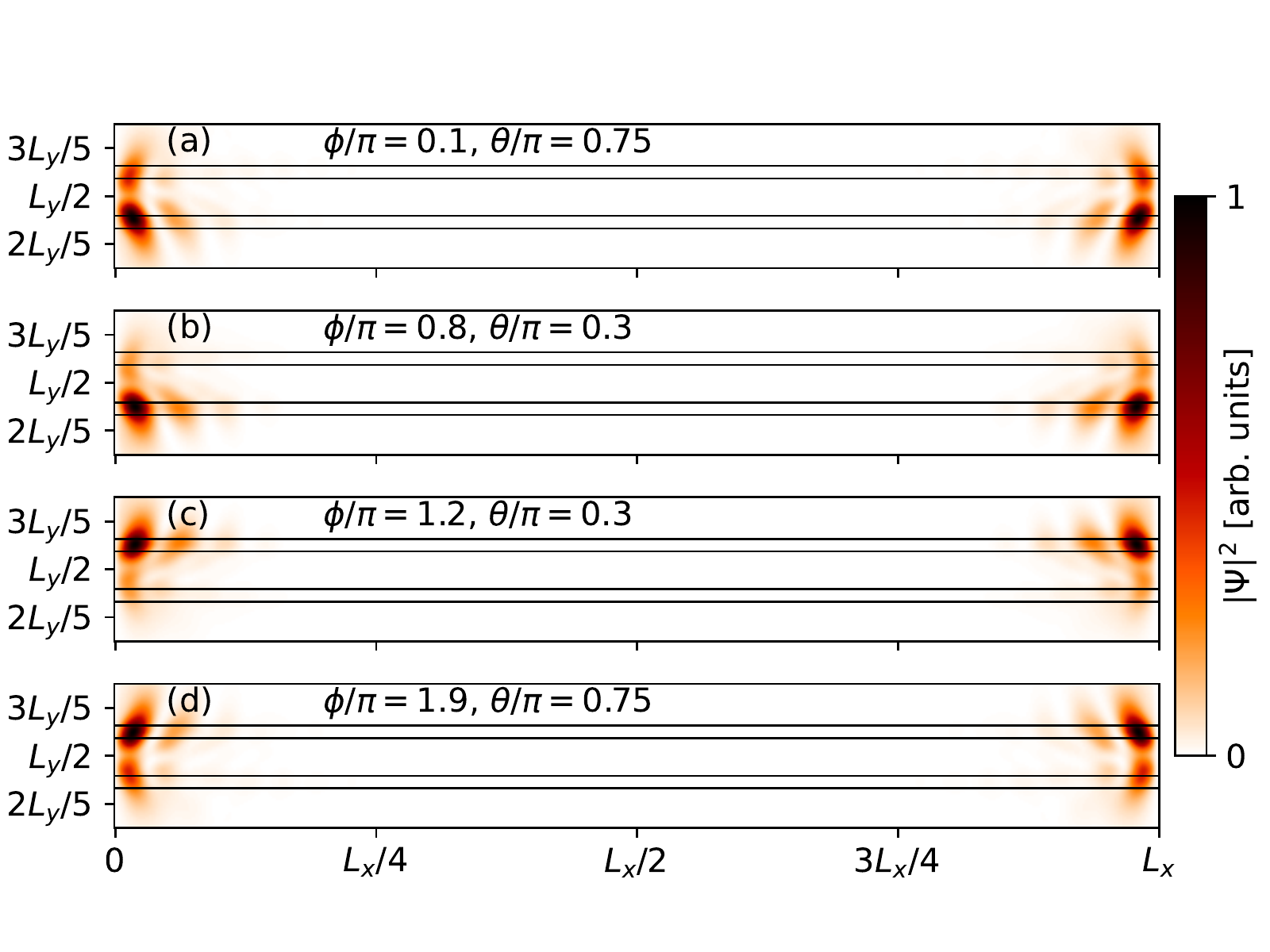}
	\caption{
		The probability density of  MBSs for different points in the phase diagram. All parameters are as in Fig.~\ref{fig:toy_model_diagrams}(d), except the superconducting phase differences $\phi$ and $\theta$, which are indicated in the corresponding panel. We note that, depending on these phases, the MBSs have larger support in one or the other junction, breaking the symmetry between them.
	}
	\label{fig:wf_full_2d}
\end{figure}

In Fig.~\ref{fig:toy_model_diagrams}(d), we show a profile cut through an MBS probability density. We show the full two-dimensional  MBS probability density in Fig.~\ref{fig:wf_full_2d}. The MBSs are localized at opposite ends and, depending on the superconducting phase, values are more localized in one or the other junction. We note that the large portion of the MBS wave function is located under the bulk superconductor, which allows one to minimize the overlap between two MBSs. These MBSs are well localized and hardly overlap in spite of a relatively small topological gap.

\begin{figure}[!b]
	\centering
	\includegraphics[width=\linewidth]{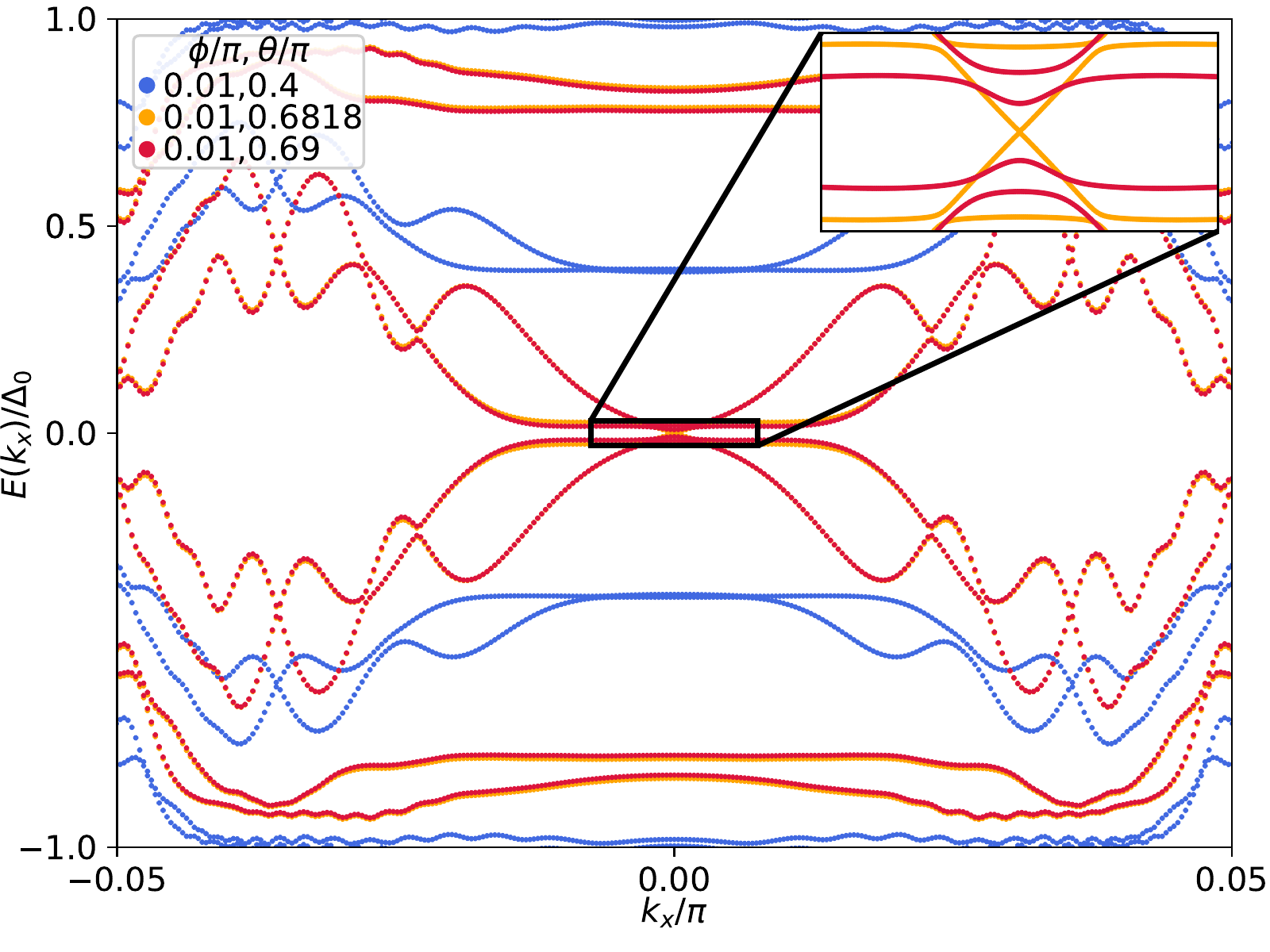}
	\caption{Energy spectrum $E(k_x)$ in the semi-infinite geometry. The parameters are for a realistic Ge system and are the same as for Fig.~\ref{fig:phase_diagram_realistic}. The data represented by the blue dots is for superconducting phase differences $\phi$ and $\theta$ deep in the trivial phase, the yellow dots are at the phase transition, and the red dots are in the topological phase.}
	\label{fig:energy_kx}
\end{figure}

\section{\label{appsec:energy_vs_kx}Energy spectrum for germanium}
In this Appendix we show the energy spectrum $E(k_x)$ in the semi-infinite geometry for realistic Ge parameters in both the trivial and topological phase, as well as at the topological phase transition; see Fig.~\ref{fig:energy_kx}. The energy spectrum shows that in the topological phase there are many in-gap states, however, as discussed in the main text, in finite length systems the trivial states are delocalized over the full length of the junction, whereas MBSs are localized at the junction ends and therefore more prominent in the LDOS.

\FloatBarrier
\nocite{adelsberger2023}
\bibliography{bibliography}

\end{document}